\documentclass[prb,twocolumn,showpacs,superscriptaddress]{revtex4}

\usepackage{bm}
\usepackage{graphicx}

\newcommand{\bOm}{\mbox{\boldmath$\Omega$}}
\newcommand{\osig}{\overline\sigma}
\newcommand{\Nabla}{\mbox{\boldmath$\nabla$}}

\renewcommand{\Im}{\mathrm{Im}\,}

\newcommand{\ba}{\begin{eqnarray}}
\newcommand{\be}{\begin{equation}}
\newcommand{\ea}{\end{eqnarray}}
\newcommand{\ee}{\end{equation}}

\newcommand{\p}{\mathbf{p}}
\newcommand{\q}{\mathbf{q}}
\newcommand{\vp}{\mathbf{v}_{\mathbf{p}}}
\newcommand{\jm}{\mathbf{j}_\mu}
\newcommand{\nm}{n_{\mathrm{i}}}
\newcommand{\sump}{\int \! \frac{d^3p}{(2\pi)^3}\,}
\newcommand{\sumpp}{\int \! \frac{d^3p'}{(2\pi)^3}\,}
\newcommand{\sumppp}{\int \! \frac{d^3p\,d^3p'}{(2\pi)^6}\,}
\newcommand{\Be}{B_{\mathrm{e}}}
\newcommand{\Bh}{B_{\mathrm{h}}}
\newcommand{\Bm}{B_{\mathrm{i}}}
\renewcommand{\ge}{g_{\mathrm{e}}}
\newcommand{\gh}{g_{\mathrm{h}}}
\newcommand{\gm}{g_{\mathrm{i}}}
\newcommand{\me}{\mu_{\mathrm{e}}}
\newcommand{\mh}{\mu_{\mathrm{h}}}
\newcommand{\mm}{\mu_{\mathrm{i}}}
\newcommand{\mub}{\mu_\mathrm{B}}
\newcommand{\tf}{\tau_{\mathrm{flip}}}
\newcommand{\ts}{\tau_{\mathrm{spin}}}
\renewcommand{\tt}{\tau_{\mathrm{tot}}}
\newcommand{\chiP}{\chi_{\mathrm{Pauli}}}
\newcommand{\chiC}{\chi_{\mathrm{Curie}}}

\begin{document}

\title{Magnetic susceptibilities of diluted magnetic semiconductors and
anomalous Hall-voltage noise}

\author{C. Timm}
\email{timm@physik.fu-berlin.de}
\affiliation{Institut f\"ur Theoretische Physik, Freie Universit\"at Berlin,
Arnimallee 14, D-14195 Berlin, Germany}
\author{F. von Oppen}
\affiliation{Institut f\"ur Theoretische Physik, Freie Universit\"at Berlin,
Arnimallee 14, D-14195 Berlin, Germany}
\author{F. H\"of\/ling}
\affiliation{Hahn-Meitner-Institut, Glienicker Stra\ss{}e 100, D-14109
Berlin, Germany}
\affiliation{Institut f\"ur Theoretische Physik, Freie Universit\"at Berlin,
Arnimallee 14, D-14195 Berlin, Germany}

\date{\today}

\begin{abstract}
The carrier spin and impurity spin densities in diluted magnetic
semiconductors are considered using a semiclassical approach. Equations of
motions for the spin densities and the carrier spin current density in the
paramagnetic phase are derived, exhibiting their coupled diffusive
dynamics. The dynamical spin susceptibilities are obtained from these
equations. The theory holds for p-type and n-type semiconductors doped with
magnetic ions of arbitrary spin quantum number. Spin-orbit coupling in the
valence band is shown to lead to anisotropic spin diffusion and to a
suppression of the Curie temperature in p-type materials. As an application
we derive the Hall-voltage noise in the paramagnetic phase. This quantity
is critically enhanced close to the Curie temperature due to the
contribution from the anomalous Hall effect.
\end{abstract}


\maketitle

\section{Introduction}
\label{sec.intro}

In recent years, a lot of progress has been made in the physics of diluted
magnetic semiconductors (DMS), in particular in III-V materials doped with
manganese. In the best studied material, (Ga,Mn)As, ferromagnetic
transition temperatures around $160\:\mathrm{K}$ have been
achieved.\cite{Schiffer,Gallagher} On the theoretical side, a Zener model
based on valence-band holes exchange-coupled to local impurity spins is
very successful in describing this material, at least in the metallic
regime.\cite{Dietal,DOM,Kreview,Dreview,web} In (Ga,Mn)As manganese acts as
an acceptor and introduces localized spins $S=5/2$ due to its half-filled
d-shell. The material is p-type but partly compensated, probably due to
arsenic antisites\cite{Potashnik,MM} and manganese
interstitials.\cite{channel} In group-IV semiconductors\cite{GeMn}
manganese plays a similar role. On the other hand, in II-VI materials
manganese introduces a spin but is isovalent with the host cations.

It has also been realized that disorder is crucial for the understanding of
the properties of DMS, even in the metallic regime.\cite{BB,JS,TSO,AD}
There are two main scattering mechanisms: disorder scattering due to the
Coulomb potential of charged donors and acceptors and spin-exchange
scattering off randomly distributed impurity spins. The Coulomb interaction
is the dominant contribution to disorder. This is due to compensation,
which leads to a lower hole concentration and thus on the one hand to the
presence of charged defects of either sign and on the other to less
effective electronic screening. Due to the large Coulomb interactions, the
defects are probably incorporated during growth in partially correlated
positions---oppositely charged donors and acceptors prefer to sit on nearby
sites---and these correlations may increase with annealing.\cite{TSO,AD} In
Ref.~\onlinecite{TSO} it was shown that equilibration of defects during
growth or annealing leads to an enormous reduction of the typical width
$(\langle V^2\rangle-\langle V\rangle^2)^{1/2}$ of the disorder potential
$V$ and to a very short correlation length of $V$, of the order of the
lattice constant. \emph{Ionic} screening is thus very effective,
whereas electronic screening is not. However, the width of the disorder
potential is still roughly of the same order as the Fermi energy so that
it cannot be neglected.

Since the correlation length is so short, a description in terms of a
\emph{delta-function correlated} disorder potential is reasonable. In this
approximation, a scattered carrier tends to loose all its momentum
information. This allows for a relatively simple description of the
scattering in the semiclassical Boltzmann approach.\cite{PN} The
spin-exchange scattering, though typically weaker than the Coulomb
scattering, is expected to become important close to the Curie temperature
$T_c$, where spin fluctuations are enhanced. A systematic study of the
effect of both types of scattering on the linear response of DMS and in
particular on transport would be desirable. For example, the resistivity
$\rho$ of (Ga,Mn)As shows a maximum or at least a shoulder at
$T_c$,\cite{Esch,Matsukura,Ohno,Hayashi,Potashnik} whereas the standard
Fisher-Langer theory\cite{FL} for fluctuation corrections to the
resistivity in ferromagnetic metals predicts an infinite \emph{derivative}
of $\rho$ at $T_c$. The origin of this weak critical behavior is that the
resistivity is dominated by scattering events with large momentum transfers
$q\sim 2k_F$, where $k_F$ is the Fermi momentum. By contrast, the magnetic
susceptibility $\chi(\q)$ of ferromagnetic metals, of Ornstein-Zernicke
form,\cite{OZ,CL} diverges only at $\q=0$.

As a step towards a comprehensive theory of disorder effects on linear
resonse and transport in DMS, we present a semiclassical theory for the
paramagnetic phase of DMS in the metallic regime. Starting from the Zener
model\cite{Dietal,DOM,Kreview,Dreview} and semiclassical Boltzmann
equations, hydrodynamic equations of motion for the carrier and impurity
spin magnetizations are derived in Sec.~\ref{sec.semi}, including Coulomb
scattering and spin-exchange scattering off magnetic impurities. Because of
the semiclassical approach, these equations hold for small momenta
$\mathbf{q}$ and frequencies $\omega$. The theory is rather general in that
it applies to both the conduction and the valence band, III-V, II-VI, and
group-IV host semiconductors, and impurities with general spin $S$. From
the equations of motion, the dynamical spin susceptibilities of carriers
and impurities are derived for small $\mathbf{q}$ and $\omega$. The
resulting semiclassical susceptibility is not of Ornstein-Zernicke form.
However, this form is presumably restored by quantum effects for $q$ of the
order of $k_F$. The semiclassical results exhibit the detailed dependence
on the various sources of scattering. We find significant differences
between the conduction-band (n-type) and valence-band (p-type) cases due to
the pronounced spin-orbit coupling in the latter. For example, spin
diffusion in the valence band is anisotropic. On the other hand, we show
that semiclassically Berry-phase effects\cite{SN,TJ} are absent from the
linear susceptibility even in the valence-band case.

It would be interesting to study the effect of spin fluctuations on the
electrical conductivity close to $T_c$ in DMS.\cite{FL} This requires the
inclusion of quantum effects at the scale of $k_F$ and thus goes
beyond the Boltzmann approach. The present theory should be a good starting
point for this generalization.

We briefly comment on related work. Sinova \textit{et al.}\cite{SJA03}
consider the damping of spin waves in the \emph{ferromagnetic} phase in the
limit $\q=0$ within a  Green-function approach. Disorder scattering is
incorporated by assuming a constant nonzero quasiparticle lifetime.
Galitski \textit{et al.}\cite{GKS03} derive the local dynamical spin
susceptibility close to $T_c$ for the strongly localized regime, opposite
to the case of weak disorder scattering considered here. In the strongly
localized case the system can be mapped onto a disordered ferromagnetic
Heisenberg model and Griffiths-McCoy singularities are important above
$T_c$.\cite{GKS03} Qi and Zhang\cite{QiZ03} consider spin diffusion in
non-magnetic materials within the Boltzmann approach. The present work goes
beyond Ref.~\onlinecite{QiZ03} in that we derive the coupled dynamics of
carrier and impurity spins in DMS, consider both conduction and valence
bands explicitly, and derive the dynamical susceptibility.

As an application we derive the fluctuations of the anomalous Hall voltage
in the \emph{paramagnetic} phase in Sec.~\ref{sec.noise}. In the absence of
an external magnetic field the average anomalous Hall voltage is zero since
the average magnetization vanishes. However, fluctuations of the
magnetization lead to nonzero Hall-voltage \emph{noise}. Three mechanisms
of the anomalous Hall effect (AHE) are discussed in the literature:
\emph{skew scattering}\cite{skew} and \emph{side-jump
scattering}\cite{side} rely on the imbalance of scattering to the right and
to the left due to spin-orbit coupling. On the other hand,
\emph{Berry-phase effects}\cite{SN} lead to an AHE in the presence of
spin-orbit coupling even without scattering. Since Jungwirth \textit{et
al.}\cite{TJ} show that the latter contribution can explain the
experimental results for DMS in the ferromagnetic phase, we also assume
this mechanism.

\section{Semiclassical theory}
\label{sec.semi}

In this section we present the semiclassical theory for the linear response
of the carrier and impurity spin magnetizations in DMS in the paramagnetic
phase. We first derive hydrodynamic equations of motion for these
magnetizations and for the carrier magnetization current. Some details are
given in App.~\ref{app.a} and \ref{app.b}. In App.~\ref{app.Berry} we
show that Berry-phase corrections are absent from the equations of motion.
Then we solve these equations to obtain the spin susceptibility. The
derivation is carried through for both the conduction and the valence band,
and for arbitrary impurity spin $S$. We use $\hbar=k_{\mathrm{B}}=1$.

\subsection{Hydrodynamic equations, conduction band}
\label{sus.hydro.con}

We start with the simpler case of conduction-band electrons
exchange-coupled to impurity spins. Spin-orbit effects can be neglected
here since the conduction band has mainly $s$-orbital character. This
description is appropriate for n-type DMS with nonzero Curie temperature.
Ferromagnetism in n-type DMS is hard to achieve due to the small exchange
interaction between electron and impurity spins and is restricted to very
low temperatures.\cite{Andre00} We assume a spherically symmetric band
$\epsilon_{\p}$ to avoid inessential complications.

We first briefly motivate the Boltzmann equations for the electron density
$n_{\p\sigma}(\mathbf{r})$, where $\sigma=\pm 1/2$ is the spin orientation,
and for the
occupation fraction $f_m$ of impurity spins with quantum number $m$ of
$S^z$. The Hamiltonian reads
\ba
\lefteqn{
H = H_{\mathrm{kin}} + J \int d^3r\, \mathbf{m}(\mathbf{r})\cdot
  \mathbf{M}(\mathbf{r}) } \nonumber \\
& & \hspace{-0.8em} {}+ \ge\mub \int d^3r\,
  \mathbf{m}(\mathbf{r})\cdot\mathbf{B}_{\mathrm{e}}^{\mathrm{ext}}
  + \gm\mub \int d^3r\,
  \mathbf{M}(\mathbf{r})\cdot\mathbf{B}_{\mathrm{i}}^{\mathrm{ext}} ,
  \hspace{1.2em}
\ea
where $\mathbf{m}$ and $\mathbf{M}$ are the electron and impurity
spin densities (oriented oppositely to the magnetizations),
respectively, averaged over microscopic volume
elements and their coupling is described by the
exchange integral $J=50\pm 5\:\mathrm{meV}\,\mathrm{nm}^3$.\cite{Ohno}
$J>0$ ($J<0$) corresponds to antiferromagnetic (ferromagnetic)
coupling. We have introduced two distinct external magnetic
fields $\mathbf{B}_{\mathrm{e}}^{\mathrm{ext}}$ and
$\mathbf{B}_{\mathrm{i}}^{\mathrm{ext}}$ acting on electron and impurity
spins, respectively, in order to obtain the linear response of each species
separately, which will prove useful in Sec.~\ref{sec.noise}.

The exchange term is decoupled at the mean-field level. We can
restrict ourselves to collinear spin configurations since the paramagnetic
susceptibility is proportional to the unit matrix in our spherical model.
We choose the magnetization direction as the $z$ axis.
The mean-field Hamiltonian of the electrons and the impurities is then
\ba
H_{\mathrm{e}} & = & H_{\mathrm{kin}}
  + \ge\mub \int d^3r\, m(\mathbf{r})\,\Be , \\
H_{\mathrm{i}} & = & \gm\mub \int d^3r\, M(\mathbf{r})\,\Bm ,
\ea
respectively. In terms of the spin magnetizations
$\me = -\ge\mub  \, \langle m\rangle$,
$\mm = -\gm\mub  \, \langle M\rangle$,
the effective fields read
\ba
\Be & = & \Be^{\mathrm{ext}} - \frac{J}{\ge\gm\mub^2}\, \mm , \\
\Bm & = & \Bm^{\mathrm{ext}} - \frac{J}{\ge\gm\mub^2}\, \me .
\ea
The single-particle energy of an electron with momentum $\p$ and spin
$\sigma=\pm1/2$ is
$E^{\mathrm{e}}_{\p\sigma} = \epsilon_{\p} + \ge\mub  \sigma \Be$.
The energy of an impurity spin with magnetic quantum number
$m$ is $E^{\mathrm{i}}_m = \gm\mub  m \Bm$.
In the absence of scattering, the semiclassical equation of motion for the
electron density $n_{\p\sigma}(\mathbf{r})$ is given by the Poisson bracket
\ba
\partial_t n_{\p\sigma} & = & -\{n_{\p\sigma}, E^{\mathrm{e}}_{\p\sigma}\}
  \nonumber \\
& = & (\Nabla_{\mathbf{r}}E^{\mathrm{e}}_{\p\sigma})\cdot
    (\Nabla_{\p}n_{\p\sigma})
  - (\Nabla_{\p}E^{\mathrm{e}}_{\p\sigma})\cdot
    (\Nabla_{\mathbf{r}}n_{\p\sigma}) \nonumber \\
& = & -\mathbf{F}_\sigma\cdot \Nabla_{\p}n_{\p\sigma}
  -\vp\,\Nabla_{\mathbf{r}}n_{\p\sigma}
\ea
with the spin-dependent force
$\mathbf{F}_\sigma=-\ge\mub \sigma\Nabla_{\mathbf{r}}\Be$
and the band velocity $\vp\cong\p/m_{\mathrm{cb}}$, where $m_{\mathrm{cb}}$
is the effective mass at the Fermi energy. We use the short-hand notation
$\partial_t$ for $\partial/\partial t$.
With scattering included we obtain the Boltzmann equation
\be
\big(\partial_t + \vp\cdot\Nabla_{\mathbf{r}} +
  \mathbf{F}_\sigma\cdot\Nabla_{\p}
  \big)\, n_{\p\sigma} = {\cal S}_{\p\sigma} ,
\ee
where ${\cal S}_{\p\sigma}$ represents collision integrals
describing various sources of scattering as discussed below.

For the impurity spins we define the occupation fraction of spins with
magnetic quantum number $m$ as $f_m$, where $\sum_m f_m = 1$. The
corresponding density is $\nm f_m$, where $\nm$ is the density of
magnetically active impurities. We neglect the contribution of
interestitial magnetic impurities.\cite{channel,BK,Edinter} The Boltzmann
equation for the density $\nm f_m$ is simply $\partial_t \nm f_m = {\cal
S}_m$, since the impurities are assumed to be immobile and purely local.

We now discuss the collision integrals. The simplest one describes disorder
scattering of the electrons,\cite{PN}
\ba
S^{\mathrm{dis}}_{\p\sigma} & = & \sumpp \frac{1}{N(0)\tau}\,
  \delta(\epsilon_{\p}-\epsilon_{\p'})\,
  \big[ n_{\p'\sigma}\,(1-n_{\p\sigma}) \nonumber \\
& & {}- n_{\p\sigma}\,
  (1-n_{\p'\sigma}) \big] \nonumber \\
& = & \sumpp \frac{1}{N(0)\tau}\,
  \delta(\epsilon_{\p}-\epsilon_{\p'})\,
  \big( n_{\p'\sigma} - n_{\p\sigma} \big) .\hspace{1.2em}
\ea
Here, $N(0)$ is the density of states at the Fermi energy for one spin
component and $1/\tau$ is the transport scattering rate. Note that
there is no change of spin $\sigma$.

The next contribution is spin-exchange scattering between electron and
impurity spins. For this we need the transition probabilities between spin
states. We write the spin operator of the electron (impurity) as
$\mathbf{s}$ ($\mathbf{S}$). The joint spin state is denoted by $|\sigma
m\rangle$. The matrix elements of the exchange coupling are
\ba
\lefteqn{ \langle \sigma m|\mathbf{s}\cdot\mathbf{S}|\sigma'm'\rangle }
  \nonumber \\
& & = \frac{1}{2} \delta_{\sigma,1/2} \delta_{\sigma',-1/2}
    \delta_{m+1,m'} \sqrt{S(S+1)-m(m+1)}
  \nonumber \\
& & \quad{}+ \frac{1}{2} \delta_{\sigma,-1/2} \delta_{\sigma',1/2}
    \delta_{m-1,m'} \sqrt{S(S+1)-m(m-1)} \nonumber \\
& & \quad{}+ \delta_{\sigma\sigma'} \delta_{mm'} \sigma m .
\ea
Note that only the $\p'=\p$ contributions to the $s^zS^z$ term are taken
care of by the mean-field decoupling. For $\p'\neq \p$ this term
expresses that carriers can also scatter off impurities due to the
exchange interaction without flipping the spins. The mean-field
approximation neglects the discreteness of the impurity spins, which this
scattering term restores.
The transition probabilities $P_{\sigma m,\sigma'm'}$ between the states
are given by the absolute square of the matrix elements,
\ba
\lefteqn{ P_{\sigma m,\sigma'm'} } \nonumber \\
& & = \frac{1}{4} \delta_{\sigma,1/2} \delta_{\sigma',-1/2}
    \delta_{m+1,m'}\, [S(S+1)-m(m+1)] \nonumber \\
& & \quad{}+ \frac{1}{4} \delta_{\sigma,-1/2} \delta_{\sigma',1/2}
    \delta_{m-1,m'}\, [S(S+1)-m(m-1)] \nonumber \\
& & \quad{}+ \frac{1}{4} \delta_{\sigma\sigma'} \delta_{mm'} m^2 .
\ea
The collision integral for electron-impurity spin scattering can then be
written as
\ba
\lefteqn{ {\cal S}_{\p\sigma m}^{\mathrm{spin}} =
  \sumpp \sum_{\sigma'm'} \frac{1}{N(0)\ts}\,
  \delta(\epsilon_{\p}+\ge\mub \sigma\Be } \nonumber \\
& & {}+ \gm\mub  m\Bm-\epsilon_{\p'}-\ge\mub \sigma'\Be-\gm\mub  m'\Bm)
\label{2.Sspin2}  \\
& & \hspace{-0.5em} {}\times P_{\sigma m,\sigma'm'}\,
  \big[ n_{\p'\sigma'}\, (1-n_{\p\sigma})\,
  f_{m'} - n_{\p\sigma}\,(1-n_{\p'\sigma'})\,f_m \big] \nonumber
\ea
with the spin-exchange scattering rate $1/\ts$.\cite{rem.taus}
Due to conservation of the total spin by the process expressed by
Eq.~(\ref{2.Sspin2}), the same
collision integral also appears in the Boltzmann equation for $f_m$. It is
the only scattering term we consider for the impurities.

The scattering processes expressed by ${\cal S}^{\mathrm{dis}}$ and ${\cal
S}^{\mathrm{spin}}$ are not sufficient for a reasonable thermodynamic
description, however. The reason is that both processes conserve the total
spin. Thus the homogeneous spin susceptibility would be zero. To avoid this
problem we allow relaxation of the total spin so that
the system can approach its thermal equilibrium. This relaxation is
implemented by an additional ``spin-flip'' scattering term for the
electrons. Physically, this can be due to the hyperfine interaction with
nuclear spins\cite{JSnuke} or electron-electron interaction in conjunction
with spin-orbit coupling in other bands.\cite{SJA03}
This process is expressed by
\ba
{\cal S}_{\p\sigma}^{\mathrm{flip}}
& = & \sumpp \frac{1}{N(0)\tf}\,
  \delta(\epsilon_{\p}+\ge\mub \sigma\Be \nonumber \\
& & {}-\epsilon_{\p'}-\ge\mub \osig\Be)\,
  \big( n_{\p'\osig} - n_{\p\sigma} \big) ,
\ea
where $\overline{\sigma}=-\sigma$.

The Boltzmann equation for the electrons now reads
\be
\left(\partial_t + \vp\cdot\Nabla_{\mathbf{r}} +
  \mathbf{F}_\sigma\cdot\Nabla_{\p}\right) n_{\p\sigma}(\mathbf{r})
  = {\cal S}_{\p\sigma}^{\mathrm{dis}}
  + {\cal S}_{\p\sigma}^{\mathrm{flip}}
  + \sum_m {\cal S}_{\p\sigma m}^{\mathrm{spin}}
\label{2.Bolh5}
\ee
and for the impurities
\be
\partial_t \nm f_m = \sump \sum_\sigma {\cal S}_{\p\sigma m}^{\mathrm{spin}}
  .
\label{2.Bolm5}
\ee
By summing Eq.~(\ref{2.Bolh5}) over $\p$, $\sigma$ one easily derives the
continuity equation $\partial_t\rho +
\Nabla_{\mathbf{r}}\cdot\mathbf{j} = 0$ for the electron number density
$\rho = \int d^3p/(2\pi)^3 \sum_\sigma n_{\p\sigma}$
and current den\-si\-ty
$\mathbf{j} = \int d^3p/(2\pi)^3 \sum_\sigma \vp\, n_{\p\sigma}$.
Our main goal is to derive corresponding
equations for the magnetizations
\ba
\me & = & -\ge \mub  \sump \sum_\sigma \sigma\, n_{\p\sigma} , \\
\mm & = & -\gm \mub  \nm \sum_m m\, f_m
\ea
and the electron magnetization current
\be
\jm = -\ge \mub  \sump \sum_\sigma \sigma\vp\, n_{\p\sigma} .
\ee
We start with the impurity spins.
Multiplying Eq.~(\ref{2.Bolm5}) by $m$ and summing over $m$ we obtain
\ba
-\frac{\partial_t \mm}{\gm\mub } & = & \sump \sum_{\sigma m} m\,
  {\cal S}_{\p\sigma m}^{\mathrm{spin}} \nonumber \\
& = & -\frac{S(S+1)}{3\ts}\, \frac{\me}{\ge\mub }
  + \frac{N(0)T}{2\ts}\, \frac{\mm}{\ge\mub \nm} \nonumber \\
& & {}+ \frac{N(0)\,S(S+1)}{6\ts}\, (\ge\mub \Be - \gm\mub \Bm)
  ,\hspace{1.5em}
\label{2.dtmm3}
\ea
to linear order in the effective fields and magnetizations,
\textit{cf.}\ App.~\ref{app.a}.
In the last expression we can identify the Pauli
susceptibility of free electrons with density of states $N(0)$ per spin
component and the Curie susceptibility of
non-interacting impurity spins with spin quantum number $S$ and
density $\nm$:\cite{Yosida}
\ba
\chiP & = & \frac{N(0)\ge^2\mub^2}{2} , \\
\chiC & = & \frac{S(S+1)\,\gm^2\mub^2\nm}{3T} .
\ea
Using these susceptibilities we write
\ba
\partial_t \mm & = & \frac{S(S+1)}{3\ts}\, \frac{\gm}{\ge}\,
  \left(\me  - \chiP\Be\right) \nonumber \\
& & {}- \frac{1}{2\ts}\, \frac{N(0)T}{\nm}\,
  \left(\mm - \chiC\Bm\right) .
\label{2.mainm}
\ea
The rate of change of the impurity magnetization $\mm$ thus depends linearly
on the deviations of $\me$ and $\mm$ from their respective equilibrium
values, which is quite reasonable.

Multiplying the Boltzmann equation (\ref{2.Bolh5}) by $\sigma$ and summing
over $\p$, $\sigma$ we obtain an equation of motion for the electron
magnetization,
\ba
\lefteqn{ -\frac{\partial_t \me}{\ge\mub } -
  \frac{\Nabla_{\mathbf{r}}\cdot\jm}{\ge\mub} } \nonumber \\
& & = \sump \sum_\sigma \sigma\, \bigg(
  {\cal S}_{\p\sigma}^{\mathrm{dis}}
  + {\cal S}_{\p\sigma}^{\mathrm{flip}}
  + \sum_m {\cal S}_{\p\sigma m}^{\mathrm{spin}} \bigg) ,\quad
\ea
where the force term on the left-hand side vanishes
since the integrand is
a total $\p$ gradient. The right-hand side can be evaluated similarly to the
calculation in App.~\ref{app.a} and expressed using $\chiP$ and $\chiC$,
\ba
\lefteqn{
\partial_t \me + \Nabla_{\mathbf{r}}\cdot\jm = - \left(\frac{2}{\tf}
  + \frac{S(S+1)}{3\ts}\right) \left(\me - \chiP\Be\right) }
  \nonumber \\
& & {}+ \frac{1}{2\ts}\, \frac{N(0)T}{\nm}\, \frac{\ge}{\gm}\,
  \left(\mm - \chiC\Bm\right) .\hspace{6.5em}
\label{2.hye4}
\ea
To eliminate the magnetization current $\jm$, we derive its equation of
motion by multiplying Eq.~(\ref{2.Bolh5}) by $\sigma\vp$ and summing over
$\p$ and $\sigma$,
\ba
\lefteqn{
-\frac{\partial_t\jm}{\ge\mub } + \sump \sum_\sigma \sigma\vp\,
  \left(\vp\cdot\Nabla_{\mathbf{r}}n_{\p\sigma}\right) } \nonumber \\
& & \quad{}- \sump \sum_\sigma \sigma\vp\, \ge\mub \sigma
  \left(\Nabla_{\mathbf{r}}\Be\right)
  \cdot \Nabla_{\p} n_{\p\sigma} \nonumber \\
& & \cong -\frac{v_{\mathrm{F}}^2}{3}\,
  \frac{\Nabla_{\mathbf{r}} \me}{\ge\mub }
  + \frac{\ge\mub \rho^{(0)}}{4m_{\mathrm{cb}}}\, \Nabla_{\mathbf{r}} \Be
  \nonumber \\
& & = -\frac{v_{\mathrm{F}}^2}{3}\, \frac{1}{\ge\mub }\, \Nabla_{\mathbf{r}}
  \left( \me - \chiP\Be \right) \nonumber \\
& & = \sump \sum_\sigma \sigma\vp\, \bigg(
  {\cal S}_{\p\sigma}^{\mathrm{dis}}
  + {\cal S}_{\p\sigma}^{\mathrm{flip}}
  + \sum_m {\cal S}_{\p\sigma m}^{\mathrm{spin}} \bigg) .\hspace{1.5em}
\label{2.hyj4}
\ea
The first term $-\partial_t\jm/\ge\mub $ is neglected since it only
becomes relevant for frequencies of the order of the largest scattering
rate. In the second term we have replaced $v_{\p}^\alpha v_{\p}^\beta$ in
the usual way by $\delta_{\alpha\beta} v_{\mathrm{F}}^2/3$,
where $v_{\mathrm{F}}$ is the Fermi velocity. This is valid since
$n_{\p\sigma}$ has significant $\mathbf{r}$ dependence only close to the
Fermi energy. The third term has been expanded to linear order
in the perturbation and
in the final step the equilibrium electron density has been written as
$\rho^{(0)} = 2N(0)m_{\mathrm{cb}} v_{\mathrm{F}}^2/3$ for a parabolic band.
Evaluating the integrals, we obtain
\be
\jm = - D\, \Nabla_{\mathbf{r}} \left( \me - \chiP\Be \right)
\label{2.jdiff3}
\ee
with the diffusion constant $D={v_{\mathrm{F}}^2\tt}/{3}$
and the total scattering rate
\be
\frac{1}{\tt} = \frac{1}{\tau} + \frac{1}{\tf}
  + \frac{S(S+1)}{4\ts} .
\ee
Inserting this result into Eq.~(\ref{2.hye4}) we find the equation of motion
of the electron spin magnetization,
\ba
\partial_t \me & = & -\left(\frac{2}{\tf}+\frac{S(S+1)}{3\ts}
  - D\nabla^2_{\mathbf{r}}\right)\, \left( \me - \chiP\Be \right)
  \nonumber \\
& & {}+ \frac{1}{2\ts}\, \frac{N(0)T}{\nm}\, \frac{\ge}{\gm}\,
  \left(\mm - \chiC\Bm\right) .
\label{2.maine}
\ea
We observe that also the rate of change of $\me$ is
linear in the deviations of the hole and impurity magnetization from their
equilibrium values. The result that $\partial_t \me$ vanishes in equilibrium
must hold in general, not just for a parabolic band, as expressed by the
Einstein relation.

The two equations (\ref{2.mainm}) and (\ref{2.maine}) are coupled both
explicitly and through the effective fields. They are formally solved by
Fourier transformation in space and time,
\ba
-i\omega \me & = & -\left(\frac{2}{\tf}+\frac{S(S+1)}{3\ts}
  + Dq^2\right) \left( \me - \chiP\Be \right) \nonumber \\
& & {}+ \frac{1}{2\ts}\, \frac{N(0)T}{\nm}\, \frac{\ge}{\gm}\,
  \left(\mm - \chiC\Bm\right) , \qquad
\label{2.maineF}  \\
-i\omega \mm & = & \frac{S(S+1)}{3\ts}\, \frac{\gm}{\ge}\,
  \left(\me  - \chiP\Be\right) \nonumber \\
& & {}- \frac{1}{2\ts}\, \frac{N(0)T}{\nm}\,
  \left(\mm - \chiC\Bm\right) .
\label{2.mainmF}
\ea
From these equations we can infer the mean-field Curie temperature $T_c$:
In the absence of external fields, the static, homogeneous magnetizations
satisfy $\me = -\chiP\, {J\mm}/{\ge\gm\mub^2}$ and $\mm = -\chiC\,
{J\me}/{\ge\gm\mub^2}$ with nonzero solutions at the
Curie temperature\cite{AbG62,JALM,Dietal,Kreview,JKSKM}
\be
T_c = \frac{S(S+1)}{6}\,N(0)\,J^2\,\nm .
\ee

\subsection{Hydrodynamic equations, valence band}
\label{sus.hydro.val}

We now derive hydrodynamic equations for valence-band holes
exchange-coupled to impurity spins, relevant for p-type DMS.
The case of spin quantum number
$S=5/2$ corresponds to substitutional Mn in GaAs. The main complication
here is the presence of spin-orbit coupling. We employ a 4-band
Kohn-Luttinger Hamiltonian\cite{KL,KL1,DOM} in the spherical approximation,
which is the simplest one incorporating the relevant physics. In the absence
of magnetic impurities the Hamiltonian reads \cite{TJ,GF}
\be
H = \frac{1}{2m}\, \left[\left(\gamma_1 + 5\gamma_2/2\right)\,k^2
  - 2\gamma_2\,(\mathbf{k}\cdot\mathbf{j})^2 \right]
\label{2.sphH2}
\ee
with Kohn-Luttinger parameters $\gamma_1$, $\gamma_2$ and the angular
momentum operator $\mathbf{j}$ of the holes, which in this subspace can be
written as a $4\times 4$ matrix and has the Casimir operator
$\mathbf{j}\cdot\mathbf{j} = 3/2(3/2+1)$.
Since the split-off band is neglected, this description only applies to
semiconductors with sufficiently strong spin-orbit coupling.
The eigenstates of $H$ at $\mathbf{k}$ are characterized by
the quantum number $j=\pm 1/2$, $\pm 3/2$
of $\hat\mathbf{k}\cdot\mathbf{j}$, where $\hat\mathbf{k}$
is the unit vector in the direction of $\mathbf{k}$, \textit{i.e.}, the spin
quantization direction is $\hat\mathbf{k}$.
We restrict ourselves to the heavy-hole band, which is
justified for energies close to the band edge because of the much smaller
density of states of the light-hole band.

We introduce the eigenstates $|j\rangle_{\mathbf{k}}$ of
$\hat\mathbf{k}\cdot\mathbf{j}$ with eigenvalues $j$. We denote the
spin eigenstates with respect to a \emph{fixed} quantization
axis $\hat\mathbf{z}$ by $|j\rangle$. The former can be expressed in
terms of the latter by means of a rotation in spin space,
\cite{Auerbach,rem.Au}
\be
|j\rangle_{\mathbf{k}} = e^{-ij^z\phi} e^{-ij^y\theta}\, |j\rangle ,
\label{2.spinrot2}
\ee
where $j^y$, $j^z$ are spin operators and $\theta$ and $\phi$ are the polar
angles of $\mathbf{k}$.

The states $|j\rangle_{\mathbf{k}}$ can be expressed in terms of
eigenstates of orbital angular momentum $\mathbf{l}$
(with $\mathbf{l}\cdot\mathbf{l}=2$)
and spin $\mathbf{s}$ with the help of
Clebsch-Gordon coefficients. One then easily finds that
\emph{all} $4\times4$ matrix elements of $s^z$ equal the corresponding
matrix elements of $j^z/3$. The same holds for the $x$ and $y$ components
because of symmetry so that $\mathbf{s}={\mathbf{j}}/{3}$
holds as an operator identity in the heavy-hole/light-hole
subspace.\cite{TJ} Consequently the heavy-hole states ($j=\pm 3/2$)
are eigenstates to
$\hat\mathbf{k}\cdot\mathbf{s}$ with eigenvalues $\pm1/2$. However, the
heavy holes alone do \emph{not} form a spin doublet since the matrix
elements of $s^\pm=l^\pm/3$ all vanish in the two-dimensional
heavy-hole subspace---single spin flips cannot change the total angular
momentum from $+3/2$ to $-3/2$ or \textit{vice versa}.

The band energy of the heavy holes is
$\epsilon_{\p} = (\gamma_1-2\gamma_2)\,p^2/2m$.
Together with the Zeeman energy their total energy is
\be
E^{\mathrm{hh}}_{\p j} = \epsilon_{\p}
  + \gh\mub \, \frac{j}{3}\,\cos\theta\, \Bh ,
\ee
where $\Bh$ is the effective magnetic field and $\theta$ is the polar angle
of $\p$ with respect to the field direction $\hat\mathbf{z}$.
Without scattering the equation of motion for the hole density reads
\ba
\partial_t n_{\p j} & = & -\{n_{\p j},E^{\mathrm{hh}}_{\p j}\}
  \nonumber \\
& = & \gh\mub \,\frac{j}{3}\,\cos\theta\, \Nabla_{\mathbf{r}}\Bh
  \cdot \Nabla_{\p} n_{\p j}
  - \frac{\p}{m_{\mathrm{hh}}} \cdot \Nabla_{\mathbf{r}} n_{\p j}
  \nonumber \\
& & {}+ \gh\mub \, \frac{j}{3}\, \sin\theta\, \Bh\,
  \frac{\mbox{\boldmath $\hat\theta$}}{p}
  \cdot \Nabla_{\mathbf{r}} n_{\p j} ,
\ea
where $m_{\mathrm{hh}}=m/(\gamma_1-2\gamma_2)$ is the heavy-hole
effective mass. This suggests to define the velocity as
\be
\vp = \frac{\p}{m_{\mathrm{hh}}} - \gh\mub \, \frac{j}{3}\,\sin\theta\,
  \Bh\,\frac{\mbox{\boldmath $\hat\theta$}}{p} .
\label{2.velo2}
\ee
Note that the second term is explicitly of first order. We should use this
velocity in the semiclassical equations. However,
we find the contribution from the second term to vanish to
first order. The reason is essentially that we have to evaluate all other
factors in equilibrium due to the explicit $\Bh$. This result is proved
together with the absence of Berry-phase contributions in
App.~\ref{app.Berry}. We thus drop the second term in Eq.~(\ref{2.velo2}).

We now turn to the derivation of the Boltzmann equation for the holes.
Analogously to the conduction-band case we have
\be
\left(\partial_t + \vp\cdot\Nabla_{\mathbf{r}} +
  \mathbf{F}_{\p j}\cdot\Nabla_{\p}\right) n_{\p j}(\mathbf{r})
  = {\cal S}_{\p j}^{\mathrm{dis}}
  + \sum_m {\cal S}_{\p j m}^{\mathrm{spin}}
\label{3.Bolh5}
\ee
with the force
$\mathbf{F}_{\p j}=-\gh\mub \,({j}/{3})\,\cos\theta\,\Nabla_{\mathbf{r}}\Bh$
for the holes and
\be
\partial_t \nm f_m = \sump \sum_j {\cal S}_{\p jm}^{\mathrm{spin}}
\label{3.Bolm5}
\ee
for the impurities.
The disorder scattering integral contains the matrix elements
${}_{\mathbf{k}}\langle j|j'\rangle_{\mathbf{k}'}
= \langle j| e^{ij^y\theta} e^{ij^z\phi} e^{-ij^z\phi'} e^{-ij^y\theta'}
|j'\rangle$.
The spin operators are $4\times 4$ matrices in the projected subspace.
For heavy holes, explicit evaluation gives the transition
probabilities
\be
\big|{}_{\mathbf{k}}\langle j|j'\rangle_{\mathbf{k}'}\big|^2
  = \left(\begin{array}{cc}
  \displaystyle\cos^6\frac{\alpha}{2} &
  \displaystyle\sin^6\frac{\alpha}{2} \\[2ex]
  \displaystyle\sin^6\frac{\alpha}{2} &
  \displaystyle\cos^6\frac{\alpha}{2}
  \end{array}\right)_{\!\!jj'}
\ee
where $j,j'=\pm 3/2$. Here, $\alpha$ is the
angle between the vectors $\mathbf{k}$ and $\mathbf{k}'$.
The collision integral for disorder scattering of heavy holes reads
\ba
\lefteqn{
{\cal S}^{\mathrm{dis}}_{\p j} = \sumpp \sum_{j'}
  \frac{1}{N(0)\tau}\, \delta\bigg(\epsilon_{\p} + \gh\mub \,\frac{j}{3}\,
  \cos\theta\,\Bh } \nonumber \\
& & - \epsilon_{\p'} - \gh\mub \,\frac{j'}{3}\,\cos\theta'\,
  \Bh\bigg)\,
  \big|{}_{\p}\langle j|j'\rangle_{\p'}\big|^2\,
  \big( n_{\p'j'} - n_{\p j} \big) . \nonumber \\[-1ex]
\label{3.Sdis3}
\ea
Note that for forward scattering ($\alpha\sim 0$) we get predominantly
$j'=j$, whereas for backscattering ($\alpha\sim\pi$) we find predominantly
$j'=-j$.

Due to the $\mathbf{k}$-dependent quantization axis the quantum number $j$
is not conserved even by pure disorder scattering due to the Elliott--Yafet
mechanism.\cite{EY} This scattering takes a hole of momentum $\mathbf{k}$
and quantum number $j$ into a state of momentum $\mathbf{k}'$ under
conservation of spin. However, its spin state is no longer an eigenstate at
$\mathbf{k}'$. In the semiclassical approximation it assumes possible
magnetic quantum numbers $j'$ with probabilities $|{}_{\p}\langle
j|j'\rangle_{\p'}|^2$.

For the hole-impurity spin scattering we need matrix elements of
$\mathbf{s}\cdot\mathbf{S}$. The transition probabilities are
\ba
\lefteqn{
P_{\p jm,\p' j'm'} = \big|{}_{\p}\langle j m|
  \mathbf{s}\cdot\mathbf{S} |j' m'\rangle_{\p'}\big|^2 } \nonumber \\
& & = \frac{1}{9}\, \bigg( \frac{1}{4}\,
  |{}_{\p}\langle j|j^+|j'\rangle_{\p'}|^2\, \delta_{m+1,m'}\,
  [S(S+1)-m(m+1)] \nonumber \\
& & \quad{}+ \frac{1}{4}\,
  |{}_{\p}\langle j|j^-|j'\rangle_{\p'}|^2\, \delta_{m-1,m'}\,
  [S(S+1)-m(m-1)] \nonumber \\
& & \quad{}+ |{}_{\p}\langle j|j^z|j'\rangle_{\p'}|^2\, \delta_{mm'} m^2
  \bigg)\quad
\ea
and the resulting collision integral reads
\ba
\lefteqn{ {\cal S}^{\mathrm{spin}}_{\p jm} = \sumpp \sum_{j'm'}
  \frac{1}{N(0)\ts}\, \delta\bigg(\epsilon_{\p} + \gh\mub \,\frac{j}{3}\,
  \cos\theta\,\Bh } \nonumber \\
& & {}+ \gm\mub  m \Bm - \epsilon_{\p'}
  - \gh\mub \frac{j'}{3} \cos\theta'\,\Bh - \gm\mub  m' \Bm\bigg)
  \nonumber \\
& & \hspace{-1em} {}\times  P_{\p jm,\p' j'm'}\,
  \big[ n_{\p'j'}\, (1-n_{\p j})\, f_{m'}
  - n_{\p j}\,(1-n_{\p'j'})\, f_m \big] . \nonumber \\[0ex]
\label{3.Sspin3}
\ea
Since the two collision integrals already include spin relaxation we do
not introduce an additional spin-flip term.

We now derive hydrodynamic equations for the hole and impurity spin
magnetizations. Some details of the calculations are shown in
App.~\ref{app.b}. The hole and impurity spin magnetizations are
\ba
\mh & = & -\gh \mub  \sump \sum_j \frac{j}{3}\,\cos\theta\,
  n_{\p j} ,
\label{3.muhdef1} \\
\mm & = & -\gm \mub  \nm \sum_m m\, f_m
\ea
and the hole magnetization current is
\be
\jm = -\gh \mub  \sump \sum_j \frac{j}{3}\,\cos\theta\,\vp\, n_{\p j} .
\ee
We start with the impurity spins. In analogy to the conduction-band case we
obtain
\ba
\partial_t \mm & = & \frac{S(S+1)}{18\ts}\, \frac{\gm}{\gh}\,
  \left(\mh - \frac{1}{3}\,\chiP\Bh\right) \nonumber \\
& & {}- \frac{5}{36\ts}\, \frac{N(0)T}{\nm}\,
  \left(\mm - \chiC\Bm\right) ,
\label{2.mainm2}
\ea
where we have again identified the Pauli susceptibility $\chiP =
N(0)\gh^2\mub^2/2$ and the Curie susceptibility. This result is of the
same linear form as for the conduction-band case. Note, however, the
presence of a factor of $1/3$ multiplying the Pauli susceptibility, which
is absent for the conduction band. This factor is easily understood
by calculating the static, homogeneous spin susceptibility of
heavy holes in the absence of impurities.
For the static susceptibility we can assume the holes to be in thermal
equilibrium,
\ba
n_{\p j} & = & n_F\!\left(\epsilon_{\p}-\mu+\gh\mub
  \,\frac{j}{3}\,\cos\theta\, \Bh\!\right) \nonumber \\
& \cong & n_F(\epsilon_{\p}-\mu)
  + n_F^{(1)}(\epsilon_{\p}-\mu)\,
  \gh\mub \,\frac{j}{3}\,\cos\theta\, \Bh ,\hspace{1.5em}
\ea
where $n_F^{(1)}(E)= n_F(E)\,[n_F(E)-1]$ is the derivative of the Fermi
function. To linear order in $\Bh$ we thus find
\ba
\mh & = & -\gh^2 \mub^2 \sump \sum_j \left(\frac{j}{3}\right)^{\!2}\,
  \cos^2\theta\, \Bh\, n_F^{(1)}(\epsilon_{\p}-\mu) \nonumber \\
& = & \frac{1}{3}\, \frac{N(0)\gh^2\mub^2}{2}\, \Bh
\;=\; \frac{1}{3}\, \chiP\, \Bh .
\ea
The extra factor stems from the angular integral over $\cos^2\theta$
and is thus due to the heavy holes not being simple spin-$1/2$ fermions, as
discussed above.

We also obtain the equation of motion for the hole magnetization,
\begin{widetext}
\ba
\lefteqn{
-\frac{\partial_t \mh}{\gh\mub } - \frac{\Nabla_{\mathbf{r}}\cdot\jm}
  {\gh\mub } - \sump \sum_j \frac{j}{3}\, \cos\theta\,
  \gh\mub \, \frac{j}{3}\, \cos\theta\, (\Nabla_{\mathbf{r}}
  \Bh)\cdot \Nabla_{\p}n_{\p j} } \nonumber \\
& & \cong -\frac{\partial_t \mh}{\gh\mub } -
  \frac{\Nabla_{\mathbf{r}}\cdot\jm}{\gh\mub } + \frac{1}{2}\, \gh\mub \,
  (\Nabla_{\mathbf{r}}\Bh) \cdot \sump \left(\Nabla_{\p}\cos^2\theta\right)\,
  n^{(0)}_{\p} \nonumber \\
& & = -\frac{\partial_t \mh}{\gh\mub }
  - \frac{\Nabla_{\mathbf{r}}\cdot\jm}{\gh\mub }
  = \sump \sum_j \frac{j}{3}\,\cos\theta\, \bigg(
  {\cal S}_{\p j}^{\mathrm{dis}}
  + \sum_m {\cal S}_{\p j m}^{\mathrm{spin}} \bigg) ,
\ea
to first order. Similarly to the calculation in App.~\ref{app.b},
we obtain for the right-hand side
\be
\partial_t \mh + \Nabla_{\mathbf{r}}\cdot\jm = - \left(\frac{1}{5\tau}
  + \frac{7S(S+1)}{180\ts}\right)\, \left(\mh - \frac{1}{3}\,\chiP\Bh\right)
  + \frac{1}{36\ts}\,\frac{N(0)T}{\nm}\,\frac{\gh}{\gm}\,
  \left(\mm - \chiC\Bm\right) .
\label{3.hye4}
\ee
To eliminate the magnetization current $\jm$ we consider its equation of
motion. The left-hand side is
\be
-\frac{\partial_t\jm}{\gh\mub } + \sump \sum_j \frac{j}{3}\,\cos\theta\,\vp\,
  \left(\vp\cdot\Nabla_{\mathbf{r}}n_{\p j}\right)
  - \sump \sum_j \frac{j}{3}\,\cos\theta\,\vp\,
  \gh\mub \, \frac{j}{3}\,\cos\theta\, \left(\Nabla_{\mathbf{r}}\Bh\right)
  \cdot \Nabla_{\p} n_{\p j} .
\label{3.q10}
\ee
\end{widetext}
The first term is again neglected.
In the second we have to be more careful because of the explicit angle
dependence. For the \emph{conduction} band the factor
$v_{\mathrm{F}}^2/3$ is obtained by assuming 
$n_{\p\sigma}$ to be the equilibrium distribution in a
constant Zeeman field. The integral over the direction of $\p$
is then easily performed. Since we obtain a term linear in
$\Nabla_{\mathbf{r}}\mu_{\mathrm{e}}$, corrections would be of higher order.
For the \emph{valence} band we also assume a constant Zeeman field,
leading to $n_{\p j}\cong n_{\p}^{(0)}+(j/3)\cos\theta\,\Delta n(p)$.
Thus the second term in Eq.~(\ref{3.q10}) becomes
\ba
\lefteqn{ v_{\mathrm{F}}^2 \sump \sum_j \frac{j}{3}\,\cos\theta\,\hat\p\,
  \bigg(\hat\p\cdot\Nabla_{\mathbf{r}} \frac{j}{3}\,\cos\theta\,\Delta n(p)
  \bigg) } \nonumber \\
& & = N(0)\, v_{\mathrm{F}}^2 \int d\xi\, \sum_j
  \left(\frac{j}{3}\right)^{\!2} \nonumber \\
& & \quad{}\times \left(\begin{array}{ccc}
    1/15 & 0 & 0 \\
    0 & 1/15 & 0 \\
    0 & 0 & 1/5
  \end{array}\right)\, \Nabla_{\mathbf{r}}\, \Delta n[p(\xi)] .\hspace{5em}
\ea
In the same approximation we find
$-\mh/\gh\mub  = N(0) \int d\xi\,\sum_j (j/3)^2 \Delta n[p(\xi)]/3$ so that
this term is
\be
-v_{\mathrm{F}}^2\, \left(\begin{array}{ccc}
  1/5 & 0 & 0 \\
  0 & 1/5 & 0 \\
  0 & 0 & 3/5
  \end{array}\right)\, \Nabla_{\mathbf{r}} \frac{\mh}{\gh\mub } .
\ee
The third term in Eq.~(\ref{3.q10}) is straightforward to evaluate to first
order,
\be
\frac{N(0)\gh\mub }{6}\, v_{\mathrm{F}}^2\,\left(\begin{array}{ccc}
  1/5 & 0 & 0 \\
  0 & 1/5 & 0 \\
  0 & 0 & 3/5
  \end{array}\right)\, \Nabla_{\mathbf{r}} \Bh ,
\ee
for a parabolic band. Again, the result holds in general due to
the Einstein relation.
Altogether the equation of motion for the magnetization current is
\ba
\lefteqn{
v_{\mathrm{F}}^2 \left(\begin{array}{ccc}
  1/5 & 0 & 0 \\
  0 & 1/5 & 0 \\
  0 & 0 & 3/5
  \end{array}\right)\, \Nabla_{\mathbf{r}}
  \left(\mh - \frac{1}{3}\,\chiP\,\Bh\right) } \nonumber \\
& & = \sump \sum_j \frac{j}{3}\,\cos\theta\,\vp\, \bigg(
  {\cal S}_{\p j}^{\mathrm{dis}}
  + \sum_m {\cal S}_{\p j m}^{\mathrm{spin}} \bigg) .\hspace{1.5em}
\ea
Evaluating the integrals we finally obtain
\be
\jm = -D\,\left(\begin{array}{ccc}
  3/5 & 0 & 0 \\
  0 & 3/5 & 0 \\
  0 & 0 & 9/5
  \end{array}\right)\, \Nabla_{\mathbf{r}}
  \left(\mh - \frac{1}{3}\,\chiP\Bh\right) ,
\ee
where we have introduced the diffusion constant $D=
{v_{\mathrm{F}}^2\tt}/{3}$ with the total relaxation rate $1/\tt=
1/(2\tau) + 5\,S(S+1)/(72\ts)$. The spin diffusion in the valence band is
thus \emph{anisotropic}. Compared with the result (\ref{2.jdiff3}) for the
conduction band, diffusion along the direction of the effective field is
enhanced and diffusion in the transverse directions is suppressed. The
origin of this interesting effect again lies in the momentum dependence of
the quantization axis due to spin-orbit coupling in conjunction with the
projection onto heavy holes: Consider, for example, heavy holes traveling
exactly along the \textit{x} direction. In the Hilbert subspace of these
holes all matrix elements of $s^z$ and $j^z$ vanish so that these holes
cannot carry any spin magnetization pointing in the \textit{z} direction.
For holes with momentum $\p$ pointing mostly but not fully in a transverse
direction the contribution to spin transport is still suppressed.

Inserting our result for the current into Eq.~(\ref{3.hye4}) we obtain the
equation of motion for the hole magnetization,
\ba
\lefteqn{
\partial_t \mh = - \bigg[\frac{1}{5\tau} + \frac{7S(S+1)}{180\ts} }
  \nonumber \\
& & \hspace{-0.3em} {}- D \left(\frac{3}{5}\,\frac{\partial^2}{\partial x^2}
  + \frac{3}{5}\,\frac{\partial^2}{\partial y^2}
  + \frac{9}{5}\,\frac{\partial^2}{\partial z^2}\right)\!
  \bigg] \left(\mh - \frac{1}{3}\,\chiP\Bh\right) \nonumber \\
& & \hspace{-0.8em} {}+ \frac{1}{36\ts}\,\frac{N(0)T}{\nm}\,\frac{\gh}{\gm}\,
  \left(\mm - \chiC\Bm\right) .
\label{2.mainh2}
\ea
This equation is of the same general form as for the
conduction band, the main differences being the reduced Pauli susceptibility
and the anisotropic spin diffusion.

As in the conduction-band case, Eqs.~(\ref{2.mainm2}) and (\ref{2.mainh2})
are coupled by the effective fields
$\Bh = \Bh^{\mathrm{ext}} - (J/\gh\gm\mub^2)\, \mm$ and
$\Bm = \Bm^{\mathrm{ext}} - (J/\gh\gm\mub^2)\, \mh$.
Fourier transformation yields
\ba
-i\omega \mh & = & - \left[\frac{1}{5\tau} + \frac{7S(S+1)}{180\ts}
  + D\, \frac{3q_x^2 + 3q_y^2 + 9q_z^2}{5} \right] \nonumber \\
& & \quad{}\times \left(\mh - \frac{1}{3}\,\chiP\Bh\right) \nonumber \\
& & {}+ \frac{1}{36\ts}\,\frac{N(0)T}{\nm}\,\frac{\gh}{\gm}\,
  \left(\mm - \chiC\Bm\right) ,
\label{2.mainh2F} \\
-i\omega \mm & = & \frac{S(S+1)}{18\ts}\, \frac{\gm}{\gh}\,
  \left(\mh - \frac{1}{3}\,\chiP\Bh\right) \nonumber \\
& & {}- \frac{5}{36\ts}\, \frac{N(0)T}{\nm}\,
  \left(\mm - \chiC\Bm\right) .
\label{2.mainm2F}
\ea
For $\omega=0$, $\mathbf{q}=0$ we find finite solutions at
\be
T = T_c = \frac{S(S+1)}{18}\,N(0)\,J^2\,\nm .
\ee
The Curie temperature of holes is \emph{reduced} by an extra factor of
$1/3$ compared to the conduction-band case. This factor stems from the same
factor in the Pauli susceptibility, which is ultimately due to spin-orbit
coupling, as discussed above. On the other hand, for typical host materials
the density of states is much \emph{higher} for the heavy holes than for
conduction-band electrons and the exchange integral $J$ is also much
larger, enhancing $T_c$ in p-type materials.

We have so far ignored the possible effect of Berry-phase contributions. As
noted in Sec.~\ref{sec.intro}, these have been found to be important for
the magnetic response of p-type DMS, where they contribute to the anomalous
Hall effect in the ferromagnetic phase.\cite{TJ} It is thus necessary to
check whether they contribute to the hydrodynamic equations. In
App.~\ref{app.Berry} we show that they do no contribute to linear order in
the effective field. Berry-phase contributions are expected in higher
orders, though.

\subsection{Susceptibilities}

With the help of the hydrodynamic equations we now
derive the linear response of the carrier spin and impurity spin
magnetizations to external fields coupled to these magnetizations.
It is useful to solve the general problem of the
Fourier-transformed hydrodynamic equations
\ba
-i\omega \me & = & -R_{\mathbf{q}}\, (\me-\alpha\chiP\Be) \nonumber \\
& & {}+ R_{\mathrm{\mathrm{ei}}}\, \frac{N(0)T}{\nm}\,
  \frac{\ge}{\gm}\, (\mm-\chiC\Bm) , \\
-i\omega \mm & = & R_{\mathrm{\mathrm{ie}}}\, \frac{\gm}{\ge}\,
  (\me-\alpha\chiP\Be) \nonumber \\
& & {}- R_{\mathrm{\mathrm{ii}}}\, \frac{N(0)T}{\nm}\, (\mm-\chiC\Bm) ,
\ea
where
\ba
\Be & = & \Be^{\mathrm{ext}} - \frac{J}{\ge\gm\mub^2}\, \mm , \\
\Bm & = & \Bm^{\mathrm{ext}} - \frac{J}{\ge\gm\mub^2}\, \me ,
\ea
which contain the special cases of the conduction band,
Eqs.~(\ref{2.maineF}) and (\ref{2.mainmF}), and the valence band,
Eqs.~(\ref{2.mainh2F}) and (\ref{2.mainm2F}).
For the valence band, the subscript ``$\mathrm{e}$'' should of course be
replaced by ``$\mathrm{h}$''.
Solving this system of equations we find
\be
\left({\me/\ge\mub  \atop \mm/\gm\mub }\right)
  = \frac{N(0)}{\det M}\,
  \left(\begin{array}{cc}
  A_{\mathrm{\mathrm{ee}}} & A_{\mathrm{\mathrm{ei}}} \\
  A_{\mathrm{\mathrm{ie}}} & A_{\mathrm{\mathrm{ii}}}
  \end{array}\right)\,
  \left({\ge\mub \Be^{\mathrm{ext}} \atop \gm\mub \Bm^{\mathrm{ext}}
    }\right)
\ee
with the determinant of the coefficient matrix
\ba
\lefteqn{ \det M = 6\nm (i\omega)^2
  + i\omega\,\big[ 2\alpha N(0)J\nm R_{\mathrm{ie}}
  - 6\nm R_{\mathbf{q}} } \nonumber \\
& & \quad{}+ 2N(0)J\nm R_{\mathrm{\mathrm{ei}}} S(S+1)
  - 6N(0)T R_{\mathrm{ii}}\big] \\
& & {}+ N(0)\, (R_{\mathrm{\mathrm{ei}}}R_{\mathrm{\mathrm{ie}}}
  -R_{\mathrm{\mathrm{ii}}}R_{\mathbf{q}})\,
  [\alpha N(0)J^2 \nm S(S+1) - 6T] \nonumber
\ea
and
\ba
A_{\mathrm{\mathrm{ee}}} & = & -3\alpha\,[i\omega\nm
  R_{\mathbf{q}}+N(0)T\,(R_{\mathrm{\mathrm{ei}}}R_{\mathrm{\mathrm{ie}}}
  -R_{\mathrm{\mathrm{ii}}}R_{\mathbf{q}})] ,\quad \\
A_{\mathrm{\mathrm{ei}}} & = & \alpha\nm\,
  [3i\omega R_{\mathrm{\mathrm{ie}}}+N(0)J\,
  (R_{\mathrm{\mathrm{ei}}}R_{\mathrm{\mathrm{ie}}}
  -R_{\mathrm{\mathrm{ii}}}R_{\mathbf{q}})
  \nonumber \\
& & {}\times S(S+1)] , \\
A_{\mathrm{\mathrm{ie}}} & = & \nm\,[2i\omega R_{\mathrm{\mathrm{ei}}}+\alpha
  N(0)J\,(R_{\mathrm{\mathrm{ei}}}R_{\mathrm{\mathrm{ie}}}
  -R_{\mathrm{\mathrm{ii}}}R_{\mathbf{q}})] \nonumber \\
& & {}\times S(S+1) , \\
A_{\mathrm{\mathrm{ii}}} & = & -2\nm\,
  [R_{\mathrm{\mathrm{ei}}}R_{\mathrm{\mathrm{ie}}}
  +R_{\mathrm{\mathrm{ii}}}\,(i\omega-R_{\mathbf{q}})]\,S(S+1) .
\ea
The magnetization becomes singular at
\be
T = T_c = \frac{\alpha}{6}\,S(S+1)\, N(0)\, J^2 \nm ,
\ee
in agreement with our earlier results.\cite{rem.Tc} We now assume
$\omega$ and $T-T_c$ to be small compared to the rates
$R_{\mathbf{q}}$, $R_{\mathrm{\mathrm{ei}}}$,
$R_{\mathrm{\mathrm{ie}}}$, $R_{\mathrm{\mathrm{ii}}}$
but do not make any assumption about $T-T_c$ vs.\ $\omega$. Then we find
\be
\left({\me \atop \mm }\right)
  = \left(\begin{array}{cc}
  \chi_{\mathrm{\mathrm{ee}}} & \chi_{\mathrm{\mathrm{ei}}} \\
  \chi_{\mathrm{\mathrm{ie}}} & \chi_{\mathrm{\mathrm{ii}}}
  \end{array}\right)\,
  \left({\Be^{\mathrm{ext}} \atop \Bm^{\mathrm{ext}}}\right)
\ee
with the susceptibility matrix
\begin{widetext}
\ba
\chi & = & \left(\begin{array}{cc}
  \chi_{\mathrm{\mathrm{ee}}} & \chi_{\mathrm{\mathrm{ei}}} \\
  \chi_{\mathrm{\mathrm{ie}}} & \chi_{\mathrm{\mathrm{ii}}}
  \end{array}\right)
\;=\; 2N(0)\, S(S+1)\, (R_{\mathrm{\mathrm{ei}}}R_{\mathrm{\mathrm{ie}}}
  -R_{\mathrm{\mathrm{ii}}}R_{\mathbf{q}})\,\mub^2
  \nonumber \\
& & \quad\times \bigg( i\omega\,\big[6 R_{\mathbf{q}} - 3\alpha N(0)J
  R_{\mathrm{ie}}
  - 2N(0)J R_{\mathrm{ei}} S(S+1)
  + \alpha N^2(0)J^2 S(S+1) R_{\mathrm{ii}}\big] \\
& & {}+ \alpha N^2(0)J^2 S(S+1)\, (R_{\mathrm{ei}}R_{\mathrm{ie}}
  -R_{\mathrm{ii}}R_{\mathbf{q}})\,
  \frac{T-T_c}{T_c} \bigg)^{\!-1}
  \left(\begin{array}{cc}
    \displaystyle \ge^2\left(\frac{\alpha N(0)J}{2}\right)^{\!2} &
    \displaystyle -\ge\gm \frac{\alpha N(0)J}{2} \\
    \displaystyle -\ge\gm \frac{\alpha N(0)J}{2} &
    \displaystyle \gm^2
  \end{array}\right) .
  \nonumber
\label{3.chigen3}
\ea
\end{widetext}
Since we cannot apply different fields to the carrier spins and the
impurity spins, the physical susceptibility of the carrier spins is
$\chi_{\mathrm{ee}}+\chi_{\mathrm{ei}}$, while the susceptibility of the
impurity spins is $\chi_{\mathrm{ie}}+\chi_{\mathrm{ii}}$. The total
susceptibility describing the response of the total magnetization is
\be
\chi_{\mathrm{tot}} = \chi_{\mathrm{ee}} + \chi_{\mathrm{ei}}
  + \chi_{\mathrm{ie}} + \chi_{\mathrm{ii}} .
\ee
Note that this physical susceptibility is always pa\-ra\-mag\-ne\-tic
since the components of the matrix factor in Eq.~(\ref{3.chigen3}) combine
to $(\ge\alpha N(0)J/2-\gm)^2$. In the static case $\omega=0$ all four
components are of Curie form.
We already see that the dimensionless parameter $-\alpha N(0)J/2$ has a
special meaning: It is the ratio between the average electron spin and the
average impurity spin in an applied field, regardless of whether the field
acts only on the electrons, on the impurities, or on both.

We now consider the special case of \emph{con\-duc\-tion-band}
\emph{elec\-trons}.
Inserting the appropriate factors from Eqs.~(\ref{2.maineF}) and
(\ref{2.mainmF}), we obtain the susceptibility matrix
\ba
\chi & = & N(0)\,S(S+1)\,\mub^2 \nonumber \\
& & {}\times \bigg(-i\omega \left[6\ts + 2S(S+1)\,\frac{(1-N(0)J/2)^2}
  {2/\tf+Dq^2}\right] \nonumber \\
& & \quad{}+ 2S(S+1)\,\left(\frac{N(0)J}{2}\right)^{\!2}\,
  \frac{T-T_c}{T_c}\bigg)^{\!-1} \nonumber \\
& & {}\times \left(\begin{array}{cc}
    \displaystyle \ge^2\left(\frac{N(0)J}{2}\right)^{\!2} &
    \displaystyle -\ge\gm \frac{N(0)J}{2} \\
    \displaystyle -\ge\gm \frac{N(0)J}{2} &
    \displaystyle \gm^2
  \end{array}\right)
\ea
with $T_c=[S(S+1)/6]\,N(0)\,J^2\nm$.
This susceptibility describes the linear response of an
n-type DMS. The same result would be
obtained for a simple model of spin-$1/2$ \emph{holes}, which is sometimes
employed in the literature.

Note that the only $\mathbf{q}$ dependence appears in the coefficient of
$\omega$. This is quite different from the standard Ornstein-Zernicke
form\cite{OZ,CL} of the susceptibility. We discuss this point further
below. The only typical length scale in $\chi$ is $\xi_{\mathrm{e}} =
\sqrt{{D\tf}/{2}}$. This is the relaxation length of the total spin since
the total spin relaxes with the spin-flip rate $1/\tf$. In the
semiclassical approximation $\xi_{\mathrm{e}}$ does not show any critical
behavior at $T_c$.

We show below that $N(0)J\ll 1$ for the valence band. For the conduction
band the density of states $N(0)$ is smaller than for the valence band and
the exchange integral $J$, which for the conduction band is predominantly
due to onsite Coulomb exchange, is also smaller so that $N(0)J$ would be
very small for n-type DMS.

The susceptibility also describes the magnetic excitations. Their
dispersion is obtained by equating the
denominator to zero and solving for $\omega$. We see that these modes are
diffusive with relaxation rates
\be
\lambda = i\omega
  = \frac{\displaystyle 2S(S+1)\,\left(\frac{N(0)J}{2}\right)^{\!2}}
  {\displaystyle 6\ts + 2S(S+1)\,\frac{(1-N(0)J/2)^2}
  {2/\tf+Dq^2}}\, \frac{T-T_c}{T_c} .
\ee
The rate $\lambda$ is always positive for $T>T_c$, as required for
exponentially decaying excitations, and is smallest for $\mathbf{q}=0$. The
$\mathbf{q}$ dependence is controlled by the total-spin relaxation length
$\xi_{\mathrm{e}}$. In the semiclassical approximation
$\lambda$ goes to zero for $T\to T_c$ for all $\mathbf{q}$ simultaneously,
but see the discussion below.

We now consider the case of \emph{valence-band holes}. Inserting the
appropriate parameter values from Eqs.~(\ref{2.mainh2F}) and
(\ref{2.mainm2F}) we obtain
\ba
\chi & = & \frac{5}{18}\, S(S+1)\, N(0)\, \mub^2 \nonumber \\
& & {}\times \bigg( -i\omega \bigg[6\ts+\frac{S(S+1)}{15}\,
  \frac{(1-5N(0)J/6)^2}{\tilde R_{\mathbf{q}}}\bigg] \nonumber \\
& & \quad{}+ \frac{5}{3}\, S(S+1)\,\left(\frac{N(0)J}{6}\right)^{\!2}\,
  \frac{T-T_c}{T_c}\bigg)^{\!-1} \nonumber \\
& & {}\times \left(\begin{array}{cc}
    \displaystyle \ge^2\left(\frac{N(0)J}{6}\right)^{\!2} &
    \displaystyle -\ge\gm \frac{N(0)J}{6} \\
    \displaystyle -\ge\gm \frac{N(0)J}{6} &
    \displaystyle \gm^2
  \end{array}\right)
\label{3.sush4}
\ea
with the Curie temperature $T_c=[S(S+1)/18]\,N(0)\,J^2\nm$ and
\ba
\lefteqn{ -\frac{5}{36\ts}\, \tilde R_{\mathbf{q}}
  = R_{hi}R_{ih} - R_{\mathrm{ii}}R_{\mathbf{q}} } \nonumber \\
& & = -\frac{5}{36\ts}\, \left(\frac{1}{5\tau} + \frac{S(S+1)}{36\ts}
  + D\,\frac{3q_x^2+3q_y^2+9q_z^2}{5} \right) \nonumber \\
& & = - \frac{5}{36\ts}\, \left(\frac{2}{5\tt}\,
  + D\,\frac{3q_x^2+3q_y^2+9q_z^2}{5} \right) .
\label{3.Rtq3}
\ea
This susceptibility applies to p-type DMS. The result is of \emph{nearly}
the same form as for the conduction band. The only differences except for
simple rescaling is that not the parameter $-N(0)J/6$ itself but
$-5N(0)J/6$ appears in the $\mathbf{q}$-dependent term and that the
diffusion is anisotropic.

For both the conduction band and the valence band the susceptibilities
depend on $\q$ only through the coefficient of the frequency $\omega$.
The \emph{static} susceptibility ($\omega=0$) is thus independent
of $\mathbf{q}$ in our approximation. This would mean that the instability
appears simultaneously at all $\mathbf{q}$. The tendency of the system to
become ferromagnetic is not found within the semiclassical Boltzmann
approach since it does not incorporate physics at large momenta $q\sim
k_F$. We expect the most important effect for $q\sim k_F$ to be the
$\mathbf{q}$ dependence of the Pauli susceptibility.\cite{Yosida} Inserting
this dependence by hand, we obtain an additional term of the order of
$+q^2/k_F^2$ in the denominator, which makes the instability first appear
at $\mathbf{q}=0$, leading to ferromagnetism. A rigorous evaluation of the
susceptibility at all momenta requires a fully quantum-mechanical
calculation, \textit{e.g.}, using the quantum Boltzmann equation. We leave
this as work for the future.

One could think that a ferromagnetic interaction between the carriers
themselves introduces a new length scale and might therefore introduce
a $q^2$ term into the denominator of $\chi$. In our approach such a
ferromagnetic coupling between the carriers, say holes, leads to an
additional term in the effective field,
\be
\Bh = \Bh^{\mathrm{ext}} - \frac{J}{\gh\gm\mub^2}\, \mm
  + \frac{K}{\gh^2\mub^2}\, \mh
\ee
with $K>0$. The derivation can be carried through. We only show the
resulting susceptibility for the valence band (the conduction-band result
is analogous),
\ba
\lefteqn{
\chi = \frac{5}{18}\, S(S+1)\, N(0)\, \mub^2\;
  \bigg(\!-i\omega } \nonumber \\
& & {}\times \bigg[6\ts
  + \frac{S(S+1)}{15}\,
  \frac{(1-5N(0)J_\kappa/6)^2}{\tilde R_{\mathbf{q}}}\bigg] \nonumber \\
& & {}+ \frac{5}{3}\,
  S(S+1)\,\left(\frac{N(0)J_\kappa}{6}\right)^{\!2}\,(1-\kappa)\,
  \frac{T-T_c^\kappa}{T_c^\kappa}\bigg)^{\!-1} \nonumber \\
& & \hspace{-0.5em} {}\times \left(\begin{array}{cc}
    \displaystyle {\ge^2}
      \left(\frac{N(0)J_\kappa}{6}\right)^{\!2} &
    \displaystyle -\ge\gm \frac{N(0)J_\kappa}{6} \\
    \displaystyle -\ge\gm \frac{N(0)J_\kappa}{6} &
    \displaystyle \gm^2
  \end{array}\right)
\ea
with $\kappa= N(0)K/6$, $J_\kappa= J/(1-\kappa)$,
and\cite{Dietal,JKSKM}
\be
T_c^\kappa = \frac{S(S+1)}{18}\,\frac{N(0)\,J^2\nm}{1-\kappa} .
\ee
The Curie temperature is enhanced by the \emph{Stoner factor}
$(1-\kappa)^{-1}$. $\tilde R_{\mathbf{q}}$ is still given by
Eq.~(\ref{3.Rtq3}). If the ferromagnetic interaction becomes so large that
$N(0)K/6=1$ then the hole system orders ferromagnetically at $T=0$ even in
the absence of any impurity spins (Stoner instability\cite{Yosida}) and our
approach breaks down. However, for DMS $\kappa$ is small.\cite{Dietal} The
same result is obtained by introducing an appropriate Landau parameter
$F_0^a= -\kappa$ into Fermi liquid theory.\cite{Dietal,PN} We see that
the inclusion of a carrier-carrier ferromagnetic exchange interaction
changes the susceptibility quantitatively but not its functional form. In
particular, it does not introduce a Ornstein-Zernicke-type $q^2$ term.

Let us estimate the parameter $N(0)J/6$: For a pa\-ra\-bo\-lic band
\be
N(0) = \frac{m_{\mathrm{hh}} k_F}{2\pi^2} = \frac{m_{\mathrm{hh}}}
  {2\pi^2}\, (3\pi^2 n)^{1/3} ,
\ee
where $n$ is the carrier density. For Ga$_{1-x}$Mn$_x$As with $x=0.05$ and
$p=0.3$ holes per manganese atom we get $N(0) \approx 7.26\times
10^{-4}\:\mathrm{meV}^{-1}\,\mathrm{nm}^{-3}$. On the other
hand,\cite{Okaba,Ohno} $J\approx(50\pm 5)\,\mathrm{meV}\:\mathrm{nm}^3$
so that
\be
\frac{N(0)\,J}{6} \approx 0.0061 .
\ee
The parameter is thus \emph{small}. However, we emphasize that the
derivation is valid for general $N(0)J$. The small value explains why the
hole contribution to the magnetization is small compared to the manganese
one.\cite{Ohno,Beschoten} Also, for antiferromagnetic coupling $J$ is
positive so that the hole and impurity spin magnetizations are opposite in
sign, in agreement with experiments.\cite{Okaba,Ando,Ohno,Beschoten}

The above estimate of $N(0)$ relies on the spherical approximation and on
the omission of the light-hole band, which are not well justified at the
hole concentration used here. A realistic Slater-Koster tight-binding
description of the unperturbed valence band\cite{SK} gives a density of
states per spin direction of $1.18\times 10^{-3}\:\mathrm{eV}^{-1}
\mbox{\AA}^{-3}$. The dimensionless parameter $N(0)J/6 \approx 0.0099$ is
thus somewhat increased by assuming a realistic band structure.

Equation (\ref{3.Rtq3}) shows that the typical length scale of $\chi$ is
$\xi_{\mathrm{h}} = \sqrt{{5D\tt}/{2}}$, which corresponds to
$\xi_{\mathrm{e}}$ in the conduction-band case. The time appearing in
$\xi_{\mathrm{h}}$ should thus be the relaxation time of the total
magnetization.

The magnetic excitations are again diffusive modes. Their relaxation rates
are
\be
\lambda = i\omega =
  \frac{\displaystyle \frac{5}{3}\,
  S(S+1)\,\left(\frac{N(0)J}{6}\right)^{\!2}}
  {\displaystyle 6\ts+\frac{S(S+1)}{15}\,
  \frac{(1-5N(0)J/6)^2}{\tilde R_{\mathbf{q}}}} \,
  \frac{T-T_c}{T_c} ,
\ee
qualitatively similar to the conduction-band case.

We propose to measure the magnetic susceptibility in the paramagnetic phase
at small $\q$ and $\omega$ for various DMS. This should allow to test the
non-standard functional form of our result. In particular, such an
experiment should look for the anisotropic spin diffusion in p-type DMS.
Studying samples with similar concentrations of magnetic impurities but
different concentrations of \emph{nonmagnetic} scatterers introduced by
codoping\cite{codope} would allow to change the scattering rate $1/\tau$
while holding $1/\ts$ and the mean-field $T_c$ nearly fixed.

\section{Anomalous Hall-voltage noise}
\label{sec.noise}

In this section we apply the semiclassical theory to the derivation of the
voltage noise in the transverse direction in the paramagnetic phase. The
average anomalous Hall voltage vanishes for $T>T_c$ due to the vanishing
average magnetization. However, \emph{fluctuations} in the magnetization
are present and are in fact critically enhanced as $T_c$ is approached.
This leads to fluctuations in the anomalous Hall voltage, which we derive
in the following. Following Ref.~\onlinecite{TJ}, we consider the
Berry-phase contribution to the anomalous Hall effect for a p-type DMS.

The fluctuations in the Hall voltage are governed by the correlation
function of the effective magnetic field acting on the hole spins. This
correlation function is closely related to the impurity-impurity spin
susceptibility $\chi_{\mathrm{ii}}$ evaluated above. Typical Hall-bar
samples are much larger than the spin relaxation length $\xi_{\mathrm{h}}$.
Hence, we can restrict ourselves to the homogeneous component,
$\mathbf{q}=0$. Fluctuations with nonzero $\mathbf{q}$ cancel out in the
macroscopic voltage measurement. On the other hand, the frequency
dependence of the $\chi_{\mathrm{ii}}$ is important since $\omega$ can
become larger than $T-T_c$ close to the transition.

We describe a p-type DMS in the metallic regime by the Hamiltonian
\be
H = H_{\mathrm{kin}}
  + J\nm \mathbf{s}\cdot\mathbf{S} - e\mathbf{E}\cdot\mathbf{r} ,
\ee
where $\mathbf{s}$ is the hole spin operator, $\mathbf{S}$ is the
\emph{averaged} ($\mathbf{q}=0$) impurity spin and $\mathbf{E}$ is a
homogeneous,
static external electric field. The external magnetic field vanishes.
The kinetic Hamiltonian is
$H_{\mathrm{kin}}=(1/2m)\, [(\gamma_1 + 5\gamma_2/2)\,p^2 -
2\gamma_2\,(\p\cdot\mathbf{j})^2]$, as in Eq.~(\ref{2.sphH2}).

The anomalous Hall conductivity has been derived by Jungwirth \textit{et
al}.\cite{TJ} The derivation is similar to the
one in App.~\ref{app.Berry}. The exchange and electric-field terms are
treated as small perturbations.  The equation of motion of $\mathbf{r}$,
Eq.~(\ref{B.eom2a}), can be rewritten as\cite{TJ}
\be
\dot\mathbf{r} = \Nabla_{\p} E_{\p j} - e\mathbf{E}\times\bOm
  + 2\,\Im \langle \Nabla_{\p} u|\partial_t u\rangle
\label{4.dotr2}
\ee
with $\bOm = \Im \langle \Nabla_{\p}u| \times |\Nabla_{\p} u\rangle$
and the heavy-hole energy
\be
E_{\p j} = \frac{p^2}{2m}(\gamma_1-2\gamma_2) + \frac{J\nm}{3}\, j\,
  \hat\p \cdot \mathbf{S} - e\mathbf{E}\cdot\mathbf{r} ,
\ee
up to first order in $\mathbf{E}$ and $\mathbf{S}$.

The charge response is derived from the Boltzmann equation $(\partial_t +
\dot\mathbf{r}\cdot\Nabla_{\mathbf{r}} + \dot\p \cdot\Nabla_{\p}) n_{\p j}
= {\cal S}^{\mathrm{dis}}_{\p j}$. We restrict ourselves to nonmagnetic
disorder scattering, assuming the disorder scattering rate $1/\tau$ to be
large compared to the spin scattering rate $1/\ts$, since inclusion of the
latter would only complicate the notation without introducing new physics.
For the anomalous Hall effect we are concerned with the \emph{charge}
density $\rho=e \int d^3p/(2\pi)^3 \sum_j n_{\p j}$ and current density
$\mathbf{j}=e \int d^3p/(2\pi)^3 \sum_j \mathbf{v}_{\p j}\, n_{\p j}$. From
a similar evaluation as in Sec.~\ref{sus.hydro.val} we recover the Drude
conductivity to order zero in the impurity spins $\mathbf{S}$.

The first contribution to the Hall current is found at first
order in $\mathbf{S}$
\be
\mathbf{j}^{(1)} = \frac{e^2 J \nm}{4\pi^2 k_F}\,
  \mathbf{E}\times\mathbf{S}\, \left(\frac{1}{\gamma_1-2\gamma_2}
  - \frac{2}{3\gamma_2}\right) m ,
\ee
where $k_F$ is the Fermi wave number in the heavy-hole band. A homogeneous
charge distribution has been assumed to obtain this result.
In the limit of large heavy-hole/light-hole mass ratio
$m_{\mathrm{hh}}/m_{\mathrm{lh}}\gg 1$ the Kohn-Luttinger parameters
satisfy $\gamma_1-2\gamma_2\ll\gamma_2$ and we obtain the simpler result
$\mathbf{j}^{(1)} = {e^2 m_{\mathrm{hh}} J \nm}/(4\pi^2
k_F)\:\mathbf{E}\times\mathbf{S}$. Note that the first-order contribution
is purely transverse---there is no anomalous contribution to the
longitudinal resistivity to this order. The anomalous Hall current density
is, to first order, $\mathbf{j}_{\mathrm{AH}}=\sigma_{\mathrm{AH}}\,
\mathbf{E}\times\hat\mathbf{S}$ with the unit vector $\hat\mathbf{S}$ in
the impurity-spin direction and\cite{TJ}
\be
\sigma_{\mathrm{AH}} = \frac{e^2 m_{\mathrm{hh}} J \nm S}{4\pi^2 k_F} .
\ee
In the paramagnetic phase the average spin magnetization vanishes and,
therefore, the average anomalous Hall current also vanishes. However, its
fluctuations $\langle
\mathbf{j}_{\mathrm{AH}}\cdot\mathbf{j}_{\mathrm{AH}}\rangle$ do not. We
write $\mathbf{j}_{\mathrm{AH}}=\tilde\sigma\,\mathbf{E}\times\mathbf{S}$,
where $\tilde\sigma=\sigma_{\mathrm{AH}}/S$. Thus
\ba
\lefteqn{
\langle \mathbf{j}_{\mathrm{AH}}\cdot\mathbf{j}_{\mathrm{AH}}\rangle
  = \tilde\sigma^2 \big\langle (\mathbf{E}\times\mathbf{S})\,
  (\mathbf{E}\times\mathbf{S}) \big\rangle } \nonumber \\
& = & \tilde\sigma^2 \bigg( E^2 \langle \mathbf{S}\cdot\mathbf{S}\rangle
  - \sum_{\alpha\beta} E_\alpha E_\beta
  \langle S^\alpha S^\beta\rangle \bigg) .
\ea
In the paramagnetic phase this gives
\be
\langle \mathbf{j}_{\mathrm{AH}}(t)\cdot\mathbf{j}_{\mathrm{AH}}(0)\rangle
  = 2\tilde\sigma^2 E^2\, \langle S^z(t)\, S^z(0) \rangle .
\ee
The time-dependent correlation function can be expressed by the
\emph{impurity-impurity} part $\chi_{\mathrm{ii}}$ of the
susceptibility in the p-type case with the help of the
fluctuation-dissipation theorem,\cite{Yosida}
\ba
\lefteqn{ \int dt\, e^{-i\omega t} \langle S^z(\mathbf{r},t)\,
  S^z(0,0) \rangle = \frac{2\,\Im \chi^{zz}_{\mathrm{ii}}(\mathbf{r},\omega)}
  {\gm^2\mub^2\nm^2\,(1-e^{-\omega/T})} } \nonumber \\
& & \cong \frac{2T\,\Im \chi^{zz}_{\mathrm{ii}}(\mathbf{r},\omega)}
  {\gm^2\mub^2\nm^2\,\omega} ,\hspace{12.5em}
\label{4.fldi2}
\ea
where the final expression is valid for $\omega\ll T$.

\begin{figure}[htb]
\includegraphics[width=2.60in]{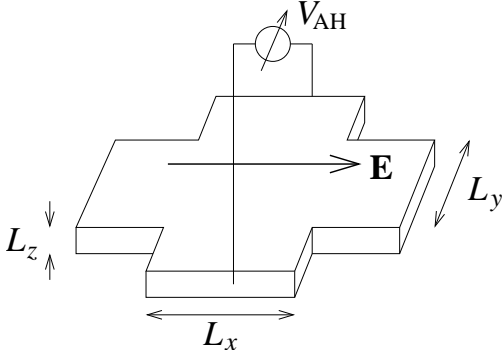}
\caption{\label{fig.LLL}Geometry of the relevant section of the Hall bar.}
\end{figure}

We now evaluate the correlation function of the anomalous Hall voltage
$U_{\mathrm{AH}}$ between the front and back sides of the relevant Hall-bar
region shown in Fig.~\ref{fig.LLL}. Since Coulomb interaction suppresses
charge fluctuations the current density is assumed to be homogeneous; 
deviations are only expected to occur at frequencies of the order of
the plasma frequency. We can then write the anomalous Hall voltage as
$U_{\mathrm{AH}}=L_y j_{\mathrm{AH}}^y/\sigma_D$, where $\sigma_D=e^2
n_{\mathrm{h}}\tau/m_{\mathrm{hh}}$ is the Drude conductivity.
If the electric field is
applied in the $x$ direction, the anomalous Hall current density in the $y$
direction is $j_{\mathrm{AH}}^y=-\tilde\sigma E S^z$. Due to homogeneity we
can average the current over the sample volume. The correlation
function of the voltage is then
\ba
\lefteqn{ \langle U_{\mathrm{AH}}(t)\, U_{\mathrm{AH}}(0) \rangle }
  \nonumber \\
& & = \frac{\tilde\sigma^2}{\sigma_D^2}\, \frac{1}{L_x^2L_z^2}\, E^2
  \int d^3r\,d^3r'\, \langle S^z(\mathbf{r},t)\, S^z(\mathbf{r}',0)\rangle
  \nonumber \\
& & = \frac{\tilde\sigma^2}{\sigma_D^2}\, \frac{L_y}{L_xL_z}\, E^2
  \int d^3r\, \langle S^z(\mathbf{r},t)\, S^z(0,0)\rangle .
\ea
Taking the Fourier transform, expressing the electric field by the voltage
applied to the relevant sample region, $E=U/L_x$,
and inserting Eq.~(\ref{4.fldi2}) we obtain
\ba
\lefteqn{ \langle U_{\mathrm{AH}} U_{\mathrm{AH}} \rangle_\omega =
  \int dt\, e^{-i\omega t} \langle U_{\mathrm{AH}}(t)\, U_{\mathrm{AH}}(0)
  \rangle } \nonumber \\
& & = \frac{\tilde\sigma^2}{\sigma_D^2}\, \frac{L_y}{L_x^3L_z}\, U^2\,
  \frac{2\,\Im\chi_{\mathrm{ii}}^{zz}(\q=0,\omega)}{\gm^2\mub^2\nm^2\,
  (1-e^{-\omega/t})} .\hspace{3em}
\ea
Assuming $1/\tau\gg 1/\ts$ and $N(0)J\ll 1$, Eq.~(\ref{3.sush4}) gives,
to leading order in $\omega$,
\be
\Im \chi_{\mathrm{ii}}^{zz}(0,\omega)
  \cong \omega\, \frac{3N(0)\,\gm^2\mub^2\, \ts}{\displaystyle
  5\,S(S+1)\,\left(\frac{N(0)J}{6}\right)^{\!4}\,
  \left(\frac{T-T_c}{T_c}\right)^{\!2}} .
\ee
For $\omega\ll T$ this leads to
\ba
\lefteqn{
\frac{\langle U_{\mathrm{AH}} U_{\mathrm{AH}} \rangle_\omega}{U^2} \cong
  \frac{\tilde\sigma^2}{\sigma_D^2}\, \frac{L_y}{L_x^3L_z}\, T }
  \nonumber \\
& & {}\times \frac{6N(0)\,\ts}{\displaystyle
  5\,S(S+1)\,\nm^2\,\left(\frac{N(0)J}{6}\right)^{\!4}\,
  \left(\frac{T-T_c}{T_c}\right)^{\!2}} \hspace{2em}
\ea
so that the noise spectrum is independent of $\omega$ for
small $\omega$. Close to the Curie temperature
the integrated noise, $\langle U_{\mathrm{AH}}^2\rangle = \langle
U_{\mathrm{AH}} U_{\mathrm{AH}} \rangle_\omega\Delta\omega$ with the
detector bandwidth $\Delta\omega=2\pi\Delta f$, satisfies
\be
\frac{\langle U_{\mathrm{AH}}^2 \rangle}{U^2} \cong
  \frac{\tilde\sigma^2}{\sigma_D^2}\, \frac{L_y}{L_x^3L_z}\,
  \frac{12\Delta\omega\,\ts}{\displaystyle 5\nm\,
  \left(\frac{N(0)J}{6}\right)^{\!2}\,
  \left(\frac{T-T_c}{T_c}\right)^{\!2}} .
\ee
The ratio of conductivities is
\be
\frac{\tilde\sigma}{\sigma_D} = \frac{3}{2}\, \frac{\nm}{n_{\mathrm{h}}}\,
  \frac{N(0)J}{6}\, \frac{1}{E_F\tau}
\ee
with the Fermi energy $E_F=k_F^2/2m_{\mathrm{hh}}$. The factor
$\nm/n_{\mathrm{h}}$ lies in the range $1\ldots 10$, the
ubiquitous factor $N(0)J/6$ drops out of the final result, and $1/E_F\tau$
has to be reasonably small for our metallic picture to apply. The final
dimensionless expression for the integrated noise is
\be
\frac{\langle U_{\mathrm{AH}}^2 \rangle}{U^2} \cong
  \frac{27}{5} \left(\frac{\nm}{n_{\mathrm{h}}}\,
  \frac{1}{E_F\tau}\right)^{\!2} \!
  \frac{L_y}{L_x^3L_z}\,\frac{1}{\nm}
  \left(\frac{T_c}{T-T_c}\right)^{\!2}
  \Delta\omega\, \ts .
\ee
This contribution to the noise is critically enhanced as the Curie
temperature is approached. In a homogeneous system it should diverge at
$T_c$ but real DMS are, by their very nature, disordered and the transition
is broadened by macroscopic inhomogeneity of $T_c$.
Furthermore, the effect strongly depends on the 
length $L_x$ of the relevant region of the Hall bar in the electric-field
direction, being large for small $L_x$. It is more weakly enhanced by a
small sample thickness $L_z$ and by a \emph{large} sample width $L_y$
across which the voltage is measured. The effect is also increased by
strong compensation ($n_{\mathrm{h}}\ll \nm$)
and in samples showing bad metallic behavior (small $E_F\tau$).

The anomalous Hall-voltage noise is in competition with the \emph{thermal}
(Johnson-Nyquist) voltage noise,\cite{Nyquist} which in integrated form is
$\langle U_{\mathrm{th}}^2\rangle = 2TR\, \Delta\omega/\pi = 2T
(L_y/\sigma_DL_xL_z) \Delta\omega/\pi$. The two contributions can be
experimentally distinguished by their different temperature and voltage
dependences. The anomalous Hall-voltage noise $\langle U_{\mathrm{AH}}^2
\rangle$ is proportional to the applied voltage squared, whereas the
thermal voltage noise is independent of voltage.

Besides being an interesting physical effect, measurement of the anomalous
Hall-voltage noise would provide an independent approach to the
impurity-spin susceptibility and to important experimental parameters, such
as the compensation fraction $n_{\mathrm{h}}/\nm$ with respect to the
density of magnetically active impurities. The Hall-voltage noise would
also provide a new way to determine the Curie temperature. More generally,
such experiments would test the applicability of the semiclassical theory
to DMS.\cite{TJ} It may also be interesting to study the anomalous
Hall-voltage noise in conventional itinerant ferromagnets such as iron.

\section{Conclusions}

A semiclassical approach based on Boltzmann equations for electrons or
holes and impurity spins has been used to derive hydrodynamic equations of
motion and spin susceptibilities of diluted magnetic semiconductors (DMS)
in the paramagnetic phase. This theory gives the leading frequency and
wave-vector dependence at small $\omega$ and $\q$. Our results apply to
p-type and n-type DMS, to III-V, II-VI, and group-IV host semiconductors,
arbitrary impurity spin quantum number $S$, and ferromagnetic or
antiferromagnetic exchange coupling $J$ of carrier and impurity spins.
While the form of the equations of motion is easy to understand, the
susceptibility has a nonstandard $\q$ dependence, which only appears in the
frequency-dependent term. Thus the semiclassical diffusive dynamics does
not lead to any $\q$ dependence of the \emph{static} susceptibility. Such
terms are expected to be introduced by physics at the much larger momentum
scale of the Fermi momentum $k_F$.

Spin-orbit coupling in the valence band leads to qualitative differences in
the susceptibility of holes compared to electrons. The first difference is
a suppression of the mean-field Curie temperature of p-type DMS compared to
n-type DMS by a factor of $1/3$, which can be traced back directly to the
momentum dependence of the spin quantization axis in the presence of
spin-orbit coupling. On the other hand, the Curie temperature in p-type DMS
is enhanced by the typically larger density of states and exchange
coupling. The second difference is the anisotropic spin diffusion in the
valence band, which is apparent in the equation of motion of the hole
magnetization and also makes the $\q$ dependence of the susceptibilities
anisotropic. The anisotropic diffusion is due to the fact that holes moving
in a direction perpendicular to the magnetization or effective field have
vanishing expectation value of the spin in the magnetization direction and
thus do not contribute to its transport.

The results have been applied to evaluate the \emph{noise} in the anomalous
Hall voltage in DMS, which is governed by the impurity-spin susceptibility
at small frequencies and momentum $\q\to 0$. Unlike the \emph{average}
anomalous Hall voltage this quantity does not vanish in the paramagnetic
phase and is even critically enhanced close to $T_c$. The noise gives an
independent experimental approach to the impurity-spin susceptibility. We
have derived the detailed dependence of the signal on the impurity and hole
concentrations and on the sample geometry.

\acknowledgments

Interesting discussions with Alex Kamenev are gratefully acknowledged. One
of us (FvO) thanks the Weizmann Institute for hospitality and support
through a Micheal Visiting Professorship while this work was completed.
This work was partially supported by the ``Junge Akademie,'' Berlin.

\appendix

\section{Hydrodynamic equations, conduction band}
\label{app.a}

In this appendix we collect a number of calculations pertaining to the
conduction-band case. The derivation of the hydrodynamic equations in
Sec.~\ref{sus.hydro.con} requires the evaluation of various integrals over
the collision terms ${\cal S}^{\mathrm{dis}}_{\p\sigma}$, ${\cal
S}^{\mathrm{spin}}_{\p\sigma m}$, and ${\cal S}^{\mathrm{flip}}_{\p\sigma}$.
We do not show all evaluations but only present a few to clarify the
method and approximations used here.

The first integral we need is
\be
\sump \sum_{\sigma m} m\, {\cal S}_{\p\sigma m}^{\mathrm{spin}} ,
\ee
which appears in the equation of motion (\ref{2.dtmm3}) of the impurity spin
magnetization. We divide the collision integral into three terms,
\be
{\cal S}_{\p\sigma m}^{\mathrm{spin}}
  = {\cal S}_{\p\sigma m}^{\mathrm{spin},0}
  + {\cal S}_{\p\sigma m}^{\mathrm{spin},+1}
  + {\cal S}_{\p\sigma m}^{\mathrm{spin},-1} ,
\ee
corresponding to $m'=m$ (no spin flip), $m'=m+1$, and $m'=m-1$,
respectively. The first contribution is
\ba
\lefteqn{ \sump \sum_{\sigma m} m\, {\cal S}_{\p\sigma m}^{\mathrm{spin},0}
  = \frac{\sum_m m^3 f_m}{4N(0)\ts} } \nonumber \\
& & {}\times \sumppp \sum_{\sigma m}
  \delta(\epsilon_{\p}-\epsilon_{\p'})\,
  \big( n_{\p'\sigma} - n_{\p\sigma} \big) = 0 ,\hspace{2em}
\label{a.sspin0.2}
\ea
as can be seen by renaming $\p\leftrightarrow\p'$ in the
term with $n_{\p\sigma}$. The other two contributions can be treated
together as
\begin{widetext}
\ba
\sump \sum_{\sigma m} m\, {\cal S}_{\p\sigma m}^{\mathrm{spin},\pm1}
& = & \frac{1}{4N(0)\ts} \sumppp \sum_m \delta(\epsilon_{\p}
  - \epsilon_{\p'} \pm\ge\mub \Be \mp\gm\mub \Bm)
  \nonumber \\
& & {}\times m\,[S(S+1)-m(m\pm 1)]\,
  \big[ n_{\p'\mp}\,(1-n_{\p\pm})\,f_{m\pm1}
  - n_{\p\pm}\,(1-n_{\p'\mp})\,f_m \big] .
\ea
\end{widetext}
We write $f_m = 1/(2S+1) + \Delta f_m$, where $\sum_m \Delta f_m=0$,
and divide the integral into terms of zero and first order in $\Delta f_m$,
\be
\sump \sum_{\sigma m} m\,{\cal S}_{\p\sigma m}^{\mathrm{spin},\pm1}
  = \Sigma^{(0)} + \Sigma^{(1)} .
\ee
In the zero-order term we expand the delta function in $\Be$, $\Bm$, and
write all terms strictly in first order. This allows to
perform the integrals,
\ba
\Sigma^{(0)} & \cong & \frac{1}{4\ts} \sum_m \frac{\mp m^2}{2S+1}
  \bigg[ \sumpp n_{\p'\mp} - \sump n_{\p\pm} \nonumber \\
& & {}+ N(0) \int d\xi\,d\xi'\, \delta^{(1)}(\xi-\xi')\,
  \big[ n_F(\xi') - n_F(\xi) \big] \nonumber \\
& & \quad{}\times (\pm\ge\mub \Be \mp\gm\mub \Bm) \bigg]
  \nonumber \\
& = & -\frac{S(S+1)}{6\ts}\, \frac{\me}{\ge\mub} \nonumber \\
& & {}+ \frac{N(0)\,S(S+1)}{12\ts}\, (\ge\mub \Be - \gm\mub \Bm) .
\ea
We have used partial integration in the last term. The
term $\Sigma^{(1)}$ in explicitly of first order in $\Delta f_m$ so that all
other factors are to be evaluated in field-free equilibrium,
\be
\Sigma^{(1)} \!= \frac{N(0)T}{4\ts} \sum_m\! m [S(S+1)-m(m\pm 1)]
  (\Delta f_{m\pm 1}-\Delta f_m) .
\label{a.q20}
\ee
In the sum we replace $m$ by $m\mp 1$ in the term containing
$f_{m\pm 1}$. If we still sum over $m$ from $-S$ to $S$, we expect
additional contributions at both ends, but these vanish due to
the factor $S(S+1)-m(m\pm 1)$. Thus we obtain
$\Sigma^{(1)}=N(0)T/4\ts\,\sum_m[-m\pm 3m^2\mp S(S+1)]\,\Delta f_m$. This
expression obviously simplifies when the contributions from ${\cal
S}^{\mathrm{spin},+1}$ and ${\cal S}^{\mathrm{spin},-1}$ are added. The
contribution from ${\cal S}^{\mathrm{spin},0}$ vanishes anyway.
Consequently, the result for the full integral is
\ba
\lefteqn{ \sump \sum_{\sigma m} m\, {\cal S}^{\mathrm{spin}}_{\p\sigma m}
  = -\frac{S(S+1)}{3\ts}\, \frac{\me}{\ge\mub } } \nonumber \\
& & \hspace{-1em} {}+ \frac{N(0)T}{2\ts}\, \frac{\mm}{\ge\mub \nm}
  + \frac{N(0) S(S+1)}{6\ts}\, (\ge\mub \Be - \gm\mub \Bm) .\nonumber
  \\[-1ex]
\ea
Note that we have expressed this result in terms of $\mm$ instead of the
occupation fractions $f_m$. This can be done in all our results so that a
closed set of equations for the two magnetizations $\me$ and $\mm$ is
obtained.

The integrals required for the equation of motion of $\me$ are quite
similar. In the equation for the magnetization current $\mathbf{j}_\mu$ we
need integrals such as
\ba
\lefteqn{
  \sump \sum_\sigma \sigma \vp\, {\cal S}_{\p\sigma}^{\mathrm{dis}} }
  \nonumber \\
& & = \frac{1}{N(0)\tau} \sumppp \sum_\sigma \sigma\vp\,\delta(\epsilon_{\p}
  - \epsilon_{\p'})\,(n_{\p'\sigma}-n_{\p\sigma}) \nonumber \\
& & = -\frac{1}{\tau} \sump \sum_\sigma \sigma\vp\, n_{\p\sigma}
  = \frac{1}{\tau}\, \frac{\mathbf{j}_\mu}{\ge\mub } ,
\ea
where the term with $n_{\p'\sigma}$ vanishes since it is odd in $\p$.
Similar evaluations are required for ${\cal S}^{\mathrm{flip}}$
and ${\cal S}^{\mathrm{spin}}$.

\section{Hydrodynamic equations, valence band}
\label{app.b}

Even though we restrict ourselves to the heavy-hole band,
the angular integrals are much more
complicated than in the conduction-band case since the explicit expression
(\ref{3.muhdef1}) for the hole magnetization $\mh$, the transition
probabilities, and the Zeeman energies now all depend on the direction in
momentum space. As noted above, the analytical expressions for the
transition probabilities are rather complicated. We use Mathematica to
analytically perform the angular integrals of the form
\be
\int \frac{d\Omega}{4\pi}\, \cos^n\theta\, \big|{}_{\p}\langle j|A
  |j'\rangle_{\p'}\big|^2
\label{b.genint2}
\ee
with $n=0,1,2$ and $A=1$, $j^z$, $j^+$, $j^-$, resulting in expressions
like
\ba
\lefteqn{
\int \frac{d\Omega}{4\pi}\, \cos^2\theta\, \big|{}_{\p}\langle j|j^\pm
  |j'\rangle_{\p'}\big|^2 } \nonumber \\
& & \hspace{-0.5em}{} = \frac{3}{40} \left\{\begin{array}{ll}
  \displaystyle \!(7-6\cos\theta'+\cos 2\theta')\,\cos^2\frac{\theta'}{2} &
  \mbox{for $j'=\mp 3/2$,} \\[1.5ex]
  \displaystyle \!(7+6\cos\theta'+\cos 2\theta')\,\sin^2\frac{\theta'}{2} &
  \mbox{for $j'=\pm 3/2$.}
  \end{array}\right. \nonumber \\[-1ex]
\ea
Here, $\theta$ and $\theta'$ are polar angles of $\p$ and $\p'$,
respectively.

As an example, we here evaluate the integral
\be
\sump \sum_{jm} m\, {\cal S}^{\mathrm{spin}}_{\p jm} ,
\ee
which corresponds to the one considered in App.~\ref{app.a}. The collision
integral is again divided into ${\cal S}^{\mathrm{spin},0}+{\cal
S}^{\mathrm{spin},+1}+{\cal S}^{\mathrm{spin},-1}$. The contribution from
${\cal S}^{\mathrm{spin},0}$ vanishes in analogy with
Eq.~(\ref{a.sspin0.2}). The other terms are expanded in $\Delta f_m$
up to linear order,
\be
\sump \sum_{jm} m\, {\cal S}^{\mathrm{spin},\pm1}_{\p jm} =
  \Sigma^{(0)}+\Sigma^{(1)} .
\ee
In $\Sigma^{(0)}$ the delta function is expanded in $\Bh$, $\Bm$ and the
term is then divided into
$\Sigma^{(0)}_{\mathrm{noflip}}+\Sigma^{(0)}_{\mathrm{flip}}$, where in the
first (second) term $j'=j$ ($j'=-j$). The first term is
evaluated similarly to the conduction-band case, taking the more complicated
angular integrals (\ref{b.genint2}) into account,
\ba
\Sigma_{\mathrm{noflip}}^{(0)}
& = & - \frac{S(S+1)}{108\ts} \bigg[ \frac{3}{2}\,\frac{\mh}{\gh\mub }
  + \frac{N(0)}{4}\, \gh\mub \Bh \nonumber \\
& & {}- \frac{5}{4}\, N(0)\, \gm\mub \Bm \bigg] .
\ea
Also, writing out $\Sigma_{\mathrm{flip}}^{(0)}$ and renaming
$j\leftrightarrow -j$ in the first term one can see that
$\Sigma_{\mathrm{flip}}^{(0)} = \Sigma_{\mathrm{noflip}}^{(0)}$.
$\Sigma^{(1)}$ can also be evaluated similarly to the conduction-band case,
\be
\Sigma^{(1)} = \frac{N(0)T}{36\ts}\, \frac{5}{2}
  \sum_m \left[ -m \pm 3m^2 \mp S(S+1) \right]\,\Delta f_m ,
\ee
which simplifies under summation over the three contributions,
\ba
\lefteqn{ \sump \sum_{jm} m\, {\cal S}^{\mathrm{spin}}_{\p jm}
= -\frac{S(S+1)}{18\ts}\, \frac{\mh}{\gh\mub } } \nonumber \\
& & {}+ \frac{N(0)\,S(S+1)}{108\ts}\, \gh\mub \Bh
  + \frac{5N(0)T}{36\ts}\, \frac{\mm}{\gm\mub \nm} \nonumber \\
& & {}- \frac{5N(0)\,S(S+1)}{108\ts}\, \gm\mub \Bm .
\ea
In the integrals pertaining to the hole magnetization and magnetization
current we obtain some terms in which the occupation fractions $f_m$ cannot
be reduced to $\mu_m$. These terms cancel in the final equations of motion
so that again a closed set of equations for $\mh$ and $\mm$ is obtained.

\section{Absence of Berry-phase contributions}
\label{app.Berry}

In the present appendix we show that Berry-phase corrections do no
contribute to the hydrodynamic equations to linear order. In the framework
of semiclassical theory they have been discussed in detail by Sundaram and
Niu.\cite{SN} If one considers a wave packet made up of electrons of a
single band, with narrow spread in real and momentum space, and with
center-of-mass position $\mathbf{r}$ and mean momentum $\p$, then the
semiclassical equations of motion for these quantities are, in the absence
of scattering,\cite{SN}
\ba
\dot\mathbf{r} & = & \Nabla_{\p}\tilde E_{\p\sigma}
  - i\,\dot p_\alpha\,
  \big(\langle\Nabla_{\p}u|\nabla_{\p}^\alpha u\rangle
    - \langle\nabla_{\p}^\alpha u|\Nabla_{\p}u\rangle\big)
  \nonumber \\
& & {}- i\,\dot r_\alpha\, \big(\langle\Nabla_{\p}u|\nabla_{\mathbf{r}}^\alpha
  u\rangle - \langle\nabla_{\mathbf{r}}^\alpha u|\Nabla_{\p}u\rangle\big)
  \nonumber \\
& & {}- i\, \big(\langle\Nabla_{\p}u|\partial_t u\rangle
  - \langle\partial_t u|\Nabla_{\p}u\rangle\big) ,
\label{B.eom2a} \\
\dot\p & = & -\Nabla_{\mathbf{r}}\tilde E_{\p\sigma}
  + i\,\dot p_\alpha\, \big(\langle\Nabla_{\mathbf{r}}u|\nabla_{\p}^\alpha u
  \rangle - \langle\nabla_{\p}^\alpha u|\Nabla_{\mathbf{r}}u\rangle\big)
  \nonumber \\
& & {}+ i\,\dot r_\alpha\,
  \big(\langle\Nabla_{\mathbf{r}}u|\nabla_{\mathbf{r}}^\alpha u\rangle
  - \langle\nabla_{\mathbf{r}}^\alpha u|\Nabla_{\mathbf{r}}u\rangle\big)
  \nonumber \\
& & {}+ i\, \big(\langle\Nabla_{\mathbf{r}}u|\partial_t u\rangle
  - \langle\partial_t u|\Nabla_{\mathbf{r}}u\rangle\big) .
\label{B.eom2b}
\ea
Summation over $\alpha=1,2,3$ is implied. $|u\rangle=|u_{\p
\sigma}\rangle$ is the periodic part of the Bloch wave function and
$\tilde E_{\p\sigma}$ is the wave-packet energy, which also contains a
Berry-phase correction,\cite{SN}
\be
\tilde E_{\p\sigma} = E_{\p\sigma} - \Im \langle \Nabla_{\mathbf{r}}
  u_{\p\sigma}|\cdot(E_{\p\sigma}-H_c)|\Nabla_{\p}u_{\p\sigma}\rangle ,
\label{B.E3}
\ee
where $H_c$ is the local Hamiltonian for the wave-packet center and
momentum and $E_{\p\sigma}$ is the corresponding eigenenergy.
This expression applies to conduction-band electrons; in the hole case
$\sigma$ should be replaced by $j$. Note that the
spatial gradient $\Nabla_{\mathbf{r}}$ acts on the \emph{center-of-mass}
vector, on which the states $|u\rangle$ depend parametrically.

For the conduction band we can immediately see that Berry-phase effects are
absent: In field-free equilibrium all spatial and temporal derivatives
vanish. The $\p$ gradients also vanish since for the Hamiltonian
$H^{(0)}=p^2/(2m_{\mathrm{cb}})$ the periodic part $|u^{(0)}\rangle$ of the
Bloch wave function is constant and the spin part $|\!\pm\!1/2\rangle$ is
also independent of $\p$. This is not changed by the Zeeman term since it
commutes with the kinetic energy in the absence of spin-orbit coupling.
Thus all terms in Eqs.~(\ref{B.eom2a}), (\ref{B.eom2b}), and (\ref{B.E3})
vanish.

For the valence band in the spherical approximation, Eq.~(\ref{2.sphH2}),
the spatial part of the Bloch wave function is also constant but the spin
part is not. It is given by Eq.~(\ref{2.spinrot2}).
The $\p$ gradient is then
\ba
\lefteqn{ \Nabla_{\p}|j\rangle_{\p} } \nonumber \\
& & = -i \bigg(j^z\,
  \frac{\mbox{\boldmath$\hat\phi$}}{p\sin\theta}\,
  e^{-ij^z\phi} e^{-ij^y\theta}
  + e^{-ij^z\phi} j^y\,\frac{\mbox{\boldmath$\hat\theta$}}{p}\,
  e^{-ij^y\theta}\bigg)\, |j\rangle . \nonumber \\[0ex]
\ea
Furthermore, the Zeeman term does not commute with the kinetic energy
so that we expect contributions from the perturbation.
We use a perturbation expansion in the effective field to obtain the
terms appearing in Eqs.~(\ref{B.eom2a}) and (\ref{B.eom2b}).
The hole Hamiltonian in the spherical approximation reads
\be
H = \underbrace{\frac{1}{2m}\,\left[(\gamma_1+5\gamma_2/2)\,p^2-2\gamma_2
  (\p\cdot\mathbf{j})^2\right]}_{H_0}
  +\underbrace{\gh\mub \, \mathbf{s}\cdot\hat\mathbf{z}\,\Bh}_{H_1} .
\ee
The unperturbed eigenenergies are
\be
\epsilon^{(0)}_{\p j} = \frac{p^2}{2m}\,
  \left\{\begin{array}{l@{\qquad}l}
    (\gamma_1-2\gamma_2) & \mbox{for $j=\pm3/2$}, \\[0.5ex]
    (\gamma_1+2\gamma_2) & \mbox{for $j=\pm1/2$}
  \end{array}\right.
\ee
and the eigenstates are $|u^{(0)}_{\p j}\rangle$,
where only the spin part has a
nontrivial $\p$ dependence.
Assuming an effective field in the $z$ direction,
the first-order perturbation is
\be
\epsilon^{(1)}_{\p j} = \gh\mub  \,\frac{1}{3}\,
  \langle j| e^{ij^y\theta} e^{ij^z\phi}\, j^z\,
  e^{-ij^z\phi} e^{-ij^y\theta} |j\rangle\, \Bh .
\ee
Restricted to heavy holes,
$\epsilon^{(1)}_{\p j} = \gh\mub\,({j}/{3})\,\cos\theta\,\Bh$.
Degenerate perturbation theory yields the perturbations to the states,
\ba
|u^{(1)}_{\p,\pm3/2}\rangle & = & \gh\mub \,\frac{1}{3}\,\Bh\,
  \Bigg( \frac{\langle u^{(0)}_{\p,1/2}|\,j^z\,|u^{(0)}_{\p,\pm3/2}\rangle}
    {\epsilon^{(0)}_{\p,\pm3/2}-\epsilon^{(0)}_{\p,1/2}}\,
    |u^{(0)}_{\p,1/2}\rangle \nonumber \\
& & {}+ \frac{\langle u^{(0)}_{\p,-1/2}|\,j^z\,|u^{(0)}_{\p,\pm3/2}\rangle}
    {\epsilon^{(0)}_{\p,\pm3/2}-\epsilon^{(0)}_{\p,-1/2}}\,
    |u^{(0)}_{\p,-1/2}\rangle \Bigg) .
\ea
Introducing the difference between heavy- and light-hole energies,
$g_{\p} = -2\gamma_2 p^2/m$, we obtain
\be
|u^{(1)}_{\p,\pm3/2}\rangle = -
  \frac{\gh\mub \,\sin\theta\,\Bh}{2\sqrt{3}\,g_{\p}}\,
  |u^{(0)}_{\p,\pm1/2}\rangle .
\label{B.u1.3}
\ee
Simplifying the notation by writing only the spin part of the wave function,
this gives
\be
|u^{(1)}_{\p,\pm3/2}\rangle = -
  \frac{\gh\mub \,\sin\theta\,\Bh}{2\sqrt{3}\,g_{\p}}\,
  e^{-ij^z\phi} e^{-ij^y\theta}\, \left|\pm\frac{1}{2}\right\rangle .
\ee
The Berry-phase correction for the energy of heavy holes, given
in Eq.~(\ref{B.E3}), is, to first order,
\ba
\lefteqn{ \Delta\epsilon_{\p j} = -\Im \langle \Nabla_{\mathbf{r}}
  u^{(1)}_{\p j}|\cdot(\epsilon^{(0)}_{\p j}-H^{(0)}_c)
  |\Nabla_{\p}u^{(0)}_{\p j}\rangle } \nonumber \\
& & = \Im\, \frac{\gh\mub \,\sin\theta\,\Nabla_{\mathbf{r}}\Bh}
  {2\sqrt{3}\,g_{\p}} \cdot
  \langle u^{(0)}_{\p,j/3}| (\epsilon^{(0)}_{\p j}-H_0) |\Nabla_{\p}
    u^{(0)}_{\p j}\rangle . \nonumber \\[-0.8ex]
\ea
Using that $|u^{(0)}_{\p,j/3}\rangle$ is an eigenstate of $H_0$ we obtain
\ba
\Delta\epsilon_{\p j} & = & \frac{\gh\mub \,\sin\theta\,
  \Nabla_{\mathbf{r}}\Bh}{2\sqrt{3}} \cdot \Im\!
  \bigg[i\,\frac{\mbox{\boldmath $\hat\phi$}}{p\sin\theta}\,
  \frac{\sqrt{3}}{2}\,\sin\theta \nonumber \\
& & {}- i\, \frac{\mbox{\boldmath $\hat\theta$}}
  {p}\,\bigg(\!\pm i\,\frac{\sqrt{3}}{2}\bigg)\bigg] \nonumber \\
& = & \frac{\gh\mub }{4p}\,
  \left(\hat\mathbf{z}\times\hat\mathbf{p}\right)\cdot
  \Nabla_{\mathbf{r}}\Bh .
\ea
This correction evidently diverges for small $\p$. The origin is the
breakdown of perturbation theory as the energy difference $g_{\p}$ between
heavy and light holes goes to zero. This divergence is not crucial here
since states deep inside the Fermi sea do not contribute to the response.

The energy entering the semiclassical equations of motion is, to
first order, $\tilde E_{\p j}=\epsilon^{(0)}_{\p j}+\epsilon^{(1)}_{\p j}
+\Delta\epsilon_{\p j}$. Thus Eq.~(\ref{B.eom2b}) reads, to first order,
\ba
\dot\p & \cong & -\Nabla_{\mathbf{r}}\tilde E_{\p\sigma} \nonumber \\
& \cong & -\gh\mub \,\frac{j}{3}\,\cos\theta\, \Nabla_{\mathbf{r}}\Bh
  - \frac{\gh\mub }{4p}\, \Nabla_{\mathbf{r}}\left[
  \left(\hat\mathbf{z}\times\hat\mathbf{p}\right)\cdot
  \Nabla_{\mathbf{r}}\Bh \right] . \nonumber \\[-1ex]
\label{B.dp5}
\ea
Thus we find an additional force which is
proportional to a second derivative of the field but independent of
the spin direction, \textit{i.e.}, an \emph{orbital} contribution.
Then Eq.~(\ref{B.eom2a}) becomes, dropping subscripts $\p$, $j$,
\ba
\dot\mathbf{r} & \cong & \Nabla_{\p}\tilde E
  - i\,\dot p_\alpha\,
  \big(\langle\Nabla_{\p}u^{(0)}|\nabla_{\p}^\alpha u^{(0)}\rangle
    - \langle\nabla_{\p}^\alpha u^{(0)}|\Nabla_{\p}u^{(0)}\rangle\big)
  \nonumber \\
& & {}- i\,\nabla_{\mathbf{\p}}^\alpha \epsilon^{(0)}\,
  \big(\langle\Nabla_{\p}u^{(0)}|\nabla_{\mathbf{r}}^\alpha u
  \rangle - \langle\nabla_{\mathbf{r}}^\alpha u|\Nabla_{\p}u^{(0)}
  \rangle\big)
  \nonumber \\
& & {}- i\, \big(\langle\Nabla_{\p}u^{(0)}|\partial_t u\rangle
  - \langle\partial_t u|\Nabla_{\p}u^{(0)}\rangle\big) .
\label{B.eom4a}
\ea
The term multiplying $\dot p_\alpha$ can be evaluated explicitly and is
found to vanish for the heavy holes. Thus
\ba
\dot\mathbf{r} & \cong & \frac{\p}{m_{\mathrm{hh}}}
  - \gh\mub \, \frac{j}{3}\,\sin\theta\,
  \Bh\,\frac{\mbox{\boldmath $\hat\theta$}}{p} \nonumber \\
& & {}- \Im \Nabla_{\p} \langle \Nabla_{\mathbf{r}}
  u|\cdot(\epsilon^{(0)}_{\p j}-H_c^{(0)})
  |\Nabla_{\p}u^{(0)}\rangle
  \nonumber \\
& & {}- i\,\frac{p_\alpha}{m_{\mathrm{hh}}}\,
  \big(\langle\Nabla_{\p}u^{(0)}|\nabla_{\mathbf{r}}^\alpha u
  \rangle - \langle\nabla_{\mathbf{r}}^\alpha u|\Nabla_{\p}u^{(0)}
  \rangle\big)
  \nonumber \\
& & {}- i\, \big(\langle\Nabla_{\p}u^{(0)}|\partial_t u\rangle
  - \langle\partial_t u|\Nabla_{\p}u^{(0)}\rangle\big) ,
\label{B.dr5}
\ea
cf.\ Eq.~(\ref{2.velo2}). To first order, the Boltzmann equation
reads
\be
\partial_t n_{\p j} + \frac{\p}{m_{\mathrm{hh}}}\cdot\Nabla_{\mathbf{r}}
  n_{\p j} + \dot\p\cdot\Nabla_{\p} n^{(0)}_{\p}
  = {\cal S}^{\mathrm{dis}}_{\p j}
  + \sum_m {\cal S}^{\mathrm{spin}}_{\p j m}
\label{B.Boltz5}
\ee
since $\Nabla_{\mathbf{r}} n_{\p j}$ and $\dot\p$ are both linear
in the perturbation. Thus the correction terms in Eq.~(\ref{B.dr5}) drop
out here and the only new term on the left-hand side comes from the orbital
force in Eq.~(\ref{B.dp5}). The equation of motion of $\mh$ is obtained
by multiplying the Boltzmann equation with $-\gh\mub (j/3)\cos\theta$ and
summing over $\p$, $j$. The orbital-force term drops out
since it contains $\sum_j j=0$.

The right-hand side of Eq.~(\ref{B.Boltz5}) also has to be
multiplied with $-\gh\mub (j/3)\cos\theta$ and summed over $\p$, $j$.
The Berry-phase
correction $\Delta\epsilon_{\p j}$ to the energy appears in the delta
functions implementing energy conservation. If we evaluate the resulting
integrals by expanding this delta function as in App.~\ref{app.b}, all terms
multiplied with $\Delta\epsilon_{\p j}$ should be evaluated to order zero.
Then the only $j$, $j'$ dependence comes from the explicit factor $j/3$ and
from the transition probabilities. However, explicit evaluation in the
$4\times4$ spin space shows that
$\sum_{jj'} (j/3)\, |{}_{\p}\langle j|j'\rangle_{\p'}|^2=0$,
$\sum_{jj'} (j/3)\, |{}_{\p}\langle j|j^z|j'\rangle_{\p'}|^2=0$,
$\sum_{jj'} (j/3)\, (|{}_{\p}\langle j|j^+|j'\rangle_{\p'}|^2
+|{}_{\p}\langle j|j^-|j'\rangle_{\p'}|^2)=0$ so that
all these terms vanish. Thus there is no contribution to
the equation of motion for the hole magnetization.

The equation of motion for the magnetization \emph{current}
$\mathbf{j}_\mu$ contains an additional factor of $\dot\mathbf{r}$ in the
integrand, which should be calculated to linear order, see
Eq.~(\ref{B.dr5}). For the left-hand side we obtain
\ba
\lefteqn{
-\frac{\partial_t\mathbf{j}_\mu}{\gh\mub }
  + \sump \sum_j \frac{j}{3}\,\cos\theta\, \dot\mathbf{r}\, \bigg(
  \frac{\p}{m_{\mathrm{hh}}}\cdot\Nabla_{\mathbf{r}} n_{\p j}
  } \nonumber \\
& & {}- \gh\mub \,\frac{j}{3}\,\cos\theta\, \Nabla_{\mathbf{r}}\Bh
  \cdot \Nabla_{\p}n^{(0)}_{\p} \nonumber \\
& & {}- \frac{\gh\mub }{4p}\, \Nabla_{\mathbf{r}}\left[
  \hat\mathbf{p}\cdot(\Nabla_{\mathbf{r}}\times\mathbf{B}_{\mathrm{h}})
  \right] \cdot\Nabla_{\p} n^{(0)}_{\p} \bigg) . \hspace{4em}
\ea
Since all terms multiplied by $\dot\mathbf{r}$ are already of first order
we replace $\dot\mathbf{r}$ by $\p/m_{\mathrm{hh}}$. Then the first two
terms in the parentheses are identical to the ones calculated above and the
third vanishes due to $\sum_j j=0$. Thus the left-hand side of the equation
of motion is not changed by Berry-phase contributions. On the right-hand
side we have to multiply the collision integrals by
$(j/3)\cos\theta\,\dot\mathbf{r}$ with
$\dot\mathbf{r}\cong\p/m_{\mathrm{hh}}+\Delta\mathbf{v}$ from
Eq.~(\ref{B.dr5}). $\Delta\mathbf{v}$ contains the term from the $\p$
dependence of the Zeeman energy as well as the Berry-phase corrections. The
contribution from $\p/m_{\mathrm{hh}}$ is what we have calculated in
Sec.~\ref{sus.hydro.val} except for the additional Berry-phase correction
$\Delta\epsilon_{\p j}$ in the delta functions. This correction is
irrelevant, however, by the argument of the previous paragraph.

In the second contribution, $\Delta\mathbf{v}$ is of first order so that
the collision integrals should be evaluated to order zero. But these are of
course zero since there is no net scattering in equilibrium. Consequently,
the linear contributions to the velocity $\dot\mathbf{r}$, in particular
the Berry-phase corrections, drop out of the equation of motion for the
magnetization current $\mathbf{j}_\mu$. In conclusion, we have shown that
the hydrodynamic equations for the valence-band case are unaffected by
Berry phases to linear order. The results of Sec.~(\ref{sus.hydro.val}) are
thus correct.


\begin{thebibliography}{99}
\bibitem{rem.taus}Note that the scattering rate $1/\ts$ is proportional to
$N(0)\,J^2\nm$, as can be seen from the Dyson equation with the
lowest-order self-energy diagram for the carriers due to spin-exchange
scattering.
\bibitem{rem.Tc}Since the spin-exchange scattering rate $1/\ts$ is
proportional to\protect\cite{rem.taus} $N(0)\,J^2\nm$ we see that $T_c$ is
proportional to $1/\ts$.
\end{thebibliography}
\end{document}